\documentclass[twocolumn,aps,prb,amsmath,amssymb,showpacs,floatfix,superscriptaddress]{revtex4-1}
\usepackage{graphicx,bm,amsmath,amssymb,natbib,color,xcolor,tabularx,multirow,array}
\usepackage{hyperref}

\begin{document}
\title{Tunable proximity effects and topological superconductivity in ferromagnetic hybrid nanowires}

\author{Samuel D. Escribano}
\affiliation{Departamento de F{\'i}sica Te{\'o}rica de la Materia Condensada C5,
Condensed Matter Physics Center (IFIMAC) and Instituto Nicol\'as  Cabrera,
 Universidad Aut{\'o}noma de Madrid, E-28049 Madrid, Spain}
\author{Elsa Prada}
 \affiliation{Instituto de Ciencia de Materiales de Madrid (ICMM), Consejo Superior de Investigaciones Cient\'{i}ficas (CSIC), E-28049 Madrid, Spain}
\author{Yuval Oreg}
\affiliation{Department of Condensed Matter Physics, Weizmann Institute of Science, Rehovot, Israel 7610}
\author{Alfredo Levy Yeyati}
\affiliation{Departamento de F{\'i}sica Te{\'o}rica de la Materia Condensada C5,
Condensed Matter Physics Center (IFIMAC) and Instituto Nicol\'as  Cabrera,
 Universidad Aut{\'o}noma de Madrid, E-28049 Madrid, Spain}
  \email[Corresponding author: ]{a.l.yeyati@uam.es}

\date{\today}

\begin{abstract}
Hybrid semiconducting nanowire devices combining epitaxial superconductor and ferromagnetic insulator layers have been recently explored experimentally as an alternative platform for topological superconductivity at zero applied magnetic field. In this proof-of-principle work we show that the topological regime can be reached in actual devices depending on some geometrical constraints. To this end, we perform numerical simulations of InAs wires in which we explicitly include the superconducting Al and magnetic EuS shells, as well as the interaction with the electrostatic environment at a self-consistent mean-field level. Our calculations show that both the magnetic and the superconducting proximity effects on the nanowire can be tuned by nearby gates thanks to their ability to move the wavefunction across the wire section.
We find that the topological phase is achieved in significant portions of the phase diagram only in configurations where the Al and EuS layers overlap on some wire facet, due to the rather local direct induced spin polarization and the appearance of an extra indirect exchange field through the superconductor. While of obvious relevance for the explanation of recent experiments, tunable proximity effects are of interest in the broader field of superconducting spintronics.

\end{abstract}

\maketitle

{\it Introduction.---} 
{Engineering} topological superconductivity in hybrid superconductor/semiconductor nanostructures where Majorana zero modes may be generated and manipulated has emerged as a great challenge for condensed matter physics in the last decade \cite{Aguado:rnc17, Lutchyn:NRM18, Prada:arxiv19}. While Rashba-coupled proximitized semiconducting nanowires appears as one of the most successful platforms \cite{Lutchyn:PRL10, Oreg:PRL10}, reaching the topological regime in these devices requires applying large magnetic fields. This turns out to be not only detrimental to superconductivity, but it also imposes some constraints in the design of quantum information processing devices \cite{Plugge:NJP17, Karzig:PRB17}.

\begin{figure}[h!]
\includegraphics[width=1\columnwidth]{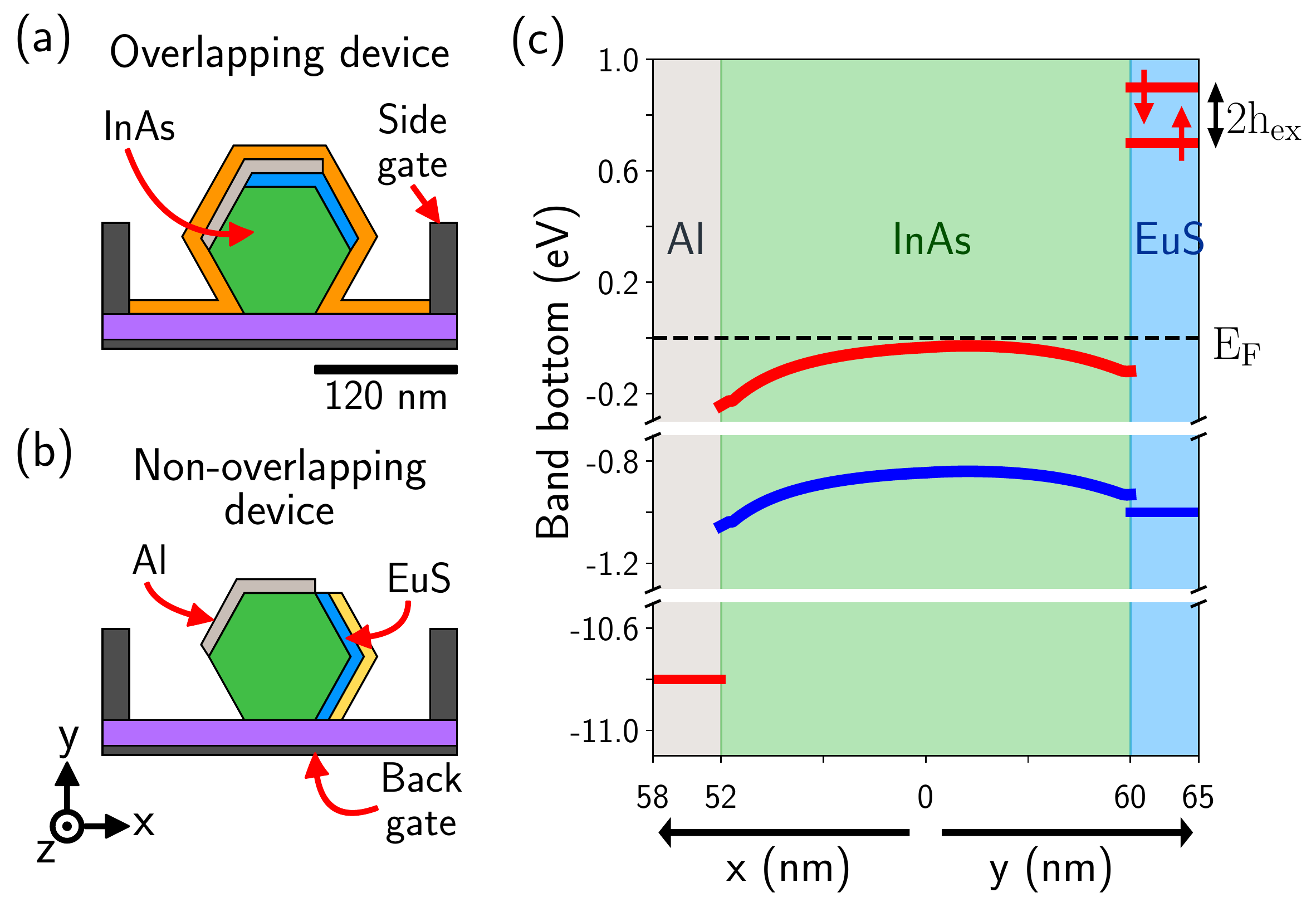}
\caption{\textbf{Hybrid nanowire geometries.} (a,b) Sketches of the devices studied in this work: a hexagonal cross section InAs nanowire (green) is simultaneously proximitized by an Al superconductor layer (light grey) and an EuS magnetic insulator layer (blue). Two side gates and one back gate (dark grey) allow to tune the chemical potential and control the position of the wavefunction inside the heterostructure. Different dielectrics are used in the experiments \cite{Vaitiekenas:NatPhys20} to allow gating (SiO$_2$, in purple) and to avoid the oxidation of the EuS layer (HfO$_2$ and AlO$_2$, in orange and yellow, respectively). {In the \textit{overlapping} device, (a), the Al and EuS layers overlap on one facet, while in the \textit{non-overlapping} device, (b), they are grown on different facets.} (c) {Diagram of the conduction and valence-band edge positions (in red and blue, respectively) across the heterostructure, spanning the three different materials. In the Al and InAs, the Fermi energy is located in the conduction band (close to the band bottom in InAs), whereas in the EuS it is inside its insulating gap. The EuS conduction band is spin splitted ({being $h_{\rm{ex}}$ the exchange coupling}). For this simulation we fix all gate voltages to zero.}}
\label{Fig1}
\end{figure}

Recent experiments \cite{Liu:AppMat20, Liu:NanoLett20, Vaitiekenas:NatPhys20} have been exploring an alternative route in which an epitaxial layer of a ferromagnetic insulator is also added to the {superconductor/semiconductor nanowire system}. While the idea of replacing the external magnetic field by the ferromagnetic layer appears as rather straightforward {in simplified models \cite{Sau:PRL10}}, there are open questions {when applied to realistic systems}. {Microscopic} calculations are required to demonstrate whether or not the topological regime could be reached for the actual geometrical and material parameters, {as well as gating conditions}. Moreover, understanding the interplay of magnetic and superconducting proximity effects in such devices is of relevance in the broader field of superconducting spintronics \cite{Wolf:Nat01, Linder:Nat15} and quantum thermodevices \cite{Giazottto:RMP06, Machon:PRL13}.

To address this problem we perform comprehensive numerical simulations of the ferromagnetic hybrid nanowire devices (see Fig. \ref{Fig1}). Related studies have been performed concurrently~\cite{Stanescu, Karsten, Wimmer, VonOppen}. We include the interaction with the electrostatic environment that typically surrounds the hybrid nanostructures by solving the Schr\"odinger-Poisson equations self-consistently in the Thomas-Fermi approximation. 

We show that topological superconductivity can {indeed} arise in these systems provided that certain geometrical and electrostatic conditions are met. We find that, for realistic values of the external gates, device layouts where the Al and EuS layers that partially cover the wire overlap on one facet develop extended topological regions in parameter space with significant topological gaps. This is in contrast to devices where the superconducting and magnetic layers are grown on adjacent facets. This could explain why recent experiments find zero bias peaks in bias spectroscopy experiments --compatible in principle with the existence of a Majorana zero mode at the wire's end --only in the former geometry but not in the latter.

Concerning the magnetization process, an open issue is whether the spin polarization is directly induced by the ferromagnet in the semiconducting nanowire electrons, or indirectly through a more elaborate process where it is first induced in the superconducting layer {(at the regions where the Al and EuS shells overlap)} and then in the wire. For instance, Ref. \onlinecite{Vaitiekenas:NatPhys20} suggested that the hysteretic behavior found in some devices could be in agreement with the indirect mechanism. We find that there is indeed an indirect induced magnetization through the Al layer, but this cannot drive a topological phase transition by itself. Conversely, there is strong direct magnetization from the EuS into the InAs, but only over a very thin region close to the ferromagnet. Interestingly, both mechanisms --direct and indirect --contribute to achieve robust and sizable topological regions in the phase diagram.

Finally, as a guide for future experiments, we elucidate the role of external potential gates in current device layouts. We show that the topological phase depends critically on the nanowire wavefunction location, a property that can be controlled by tuning appropriately those gates.

{\it Device geometries and {model}.---}
{Following closely the experiments of} Ref. \onlinecite{Vaitiekenas:NatPhys20}, we consider the two types of device geometries depicted in Fig. \ref{Fig1}(a) and (b). In both cases, a hexagonal cross-section InAs nanowire is partially covered by epitaxial Al and EuS layers. The main difference between them is that, in the \textit{overlapping} device [Fig. \ref{Fig1}(a)], the Al and EuS layers partially overlap on one facet, while in the \textit{non-overlapping} one [Fig. \ref{Fig1}(b)] they lie on adjacent facets. Various dielectrics surrounding the hybrid wires are included in our electrostatic simulations although we find that they play a minor role. Last, there are three gate electrodes used in the experiments to tune the electrostatic potential inside the devices: one back-gate and two side-gates. We analyze other geometries in the Supplemental Material (SM) \onlinecite{Supplementary}.

In this work we address the bulk electronic properties of these hybrid nanowires, which we assume are translational invariant along the $z$ direction. 
A schematic band diagram of the three different materials in the transverse directions, $x,y$, can be seen in Fig. \ref{Fig1}(c). The Al layer is a metal whose conduction band lies at $-11.7$ eV below the Fermi level \cite{AschroftMermin:76}. Despite the fact that the conduction band of the InAs is typically at the Fermi level, experimental angle-resolved photoemission spectroscopy (ARPES) \cite{Schuwalow:arxiv19} and scanning tunneling microscopy (STM) \cite{Reiner:PRX20} measurements on epitaxial Al/InAs structures show that there is a band offset of $\sim 0.2$ eV between the Al and the InAs. This imposes an electron doping of the InAs conduction band close to the Al/InAs interface. On the other hand, soft x-ray ARPES (SX-ARPES) experiments on the EuS/InAs interface \cite{Liu:AppMat20} indicate that the InAs conduction band lies well within the EuS band gap, which is of the order of $1.7$ eV \cite{Eastman:PRL69}. In particular, the EuS conduction band is located 0.7 eV above the Fermi level and the $4f$ valence bands $1$ eV below \cite{Eastman:PRL69, Liu:AppMat20}. The EuS conduction band is characterized by an exchange field $h_{\rm ex}$ that {shifts} the spin-up and spin-down energies by roughly $\pm 100$ meV \cite{Mauger:PhysRep86, Moodera:IOP07, Alphenaar:09}. In addition to this, and similarly to the InAs/Al interface, SX-ARPES experiments \cite{Liu:AppMat20} also revealed a band bending of the order of $0.1$ eV at the InAs/EuS interface, which imposes a smaller charge accumulation at this junction as well. All these band alignments are further distorted by the electric fields defined by the gate electrodes. However, for sufficiently small fields one can assume that only the InAs conduction band {moves} and can neglect its hybridization with the EuS valence bands (see the SM \cite{Supplementary} for further details).

Under these assumptions, we describe the wires using the following continuous model Hamiltonian
\begin{eqnarray}
H = \left[ \vec{k}^T  \frac{\hbar^2}{2m_{\rm eff}(\vec{r})} \vec{k} - E_{\rm F}(\vec{r}) + e\phi(\vec{r}) +h_{\rm ex}(\vec{r})\sigma_z \right]\tau_z  \nonumber\\ 
 +\frac{1}{2} \left[\vec{\alpha}(\vec{r})\cdot \left(\vec{\sigma}\times\vec{k}\right) + \left(\vec{\sigma}\times\vec{k}\right)\cdot \vec{\alpha}(\vec{r}) \right]\tau_z + \Delta(\vec{r})\sigma_y\tau_y, \ \ \
\label{hamiltonian}
\end{eqnarray}
where $\vec{r}=(x,y)$, $\vec{k}=(-i\vec{\nabla}_r,k_z)$, and $\sigma_{\alpha}$ and $\tau_{\alpha}$ denote Pauli matrices in spin and Nambu spaces, respectively. The parameters $m_{\rm eff}$, $E_{\rm F}$, $h_{\rm ex}$ and $\Delta$, corresponding to the effective mass, Fermi energy, exchange field and superconducting pairing amplitude, are taken differently for each region according to estimations from the literature, as summarized in Table I of the SM \cite{Supplementary}. To simulate the disordered outer surface of the Al layer and the irregular EuS/Al interface we introduce a random Gaussian noise in $E_F(\vec{r})$ \cite{Supplementary}.
The other parameters, i.e., the electrostatic potential $\phi(\vec{r})$ and the spin-orbit coupling (SOC) inside the wire $\alpha(\vec{r})$, are determined in a self-consistent way. For this purpose, we obtain $\phi(\vec{r})$ by solving the Schr\"odinger-Poisson equations within the Thomas-Fermi approximation \cite{Mikkelsen:PRX18, Escribano:PRB19}. The SOC $\alpha(\vec{r})$ varies locally with the electric field and is accurately calculated using the procedure described in Ref. \onlinecite{Escribano:PRR20}, see also \onlinecite{Supplementary}. Notice that the exchange field does not give rise to a magnetic orbital term in the Hamiltonian, as opposed to what happens in wires under an external magnetic field \cite{Nijholt:PRB16, Kiczek:IOP17, Dmytruk:PRB18, Winkler:PRB19}.

To obtain the electronic properties we diagonalize Eq.~\eqref{hamiltonian}. To this end, we discretize it {into a tight-binding Hamiltonian} using an appropriate mesh, which is dictated by the Al Fermi wavelength \cite{Supplementary}. Notice that a description of the three material regions (Al, InAs and EuS) on the same footing constitutes a demanding {computational} task. In the second part of this work we build a simplified model in which we integrate out the Al and EuS layers and include their {proximity effects over the InAs wire as position-dependent effective parameters}. This allows us to explore the system's phase diagram for a broader range of parameters.

\begin{figure}
\includegraphics[width=1\columnwidth]{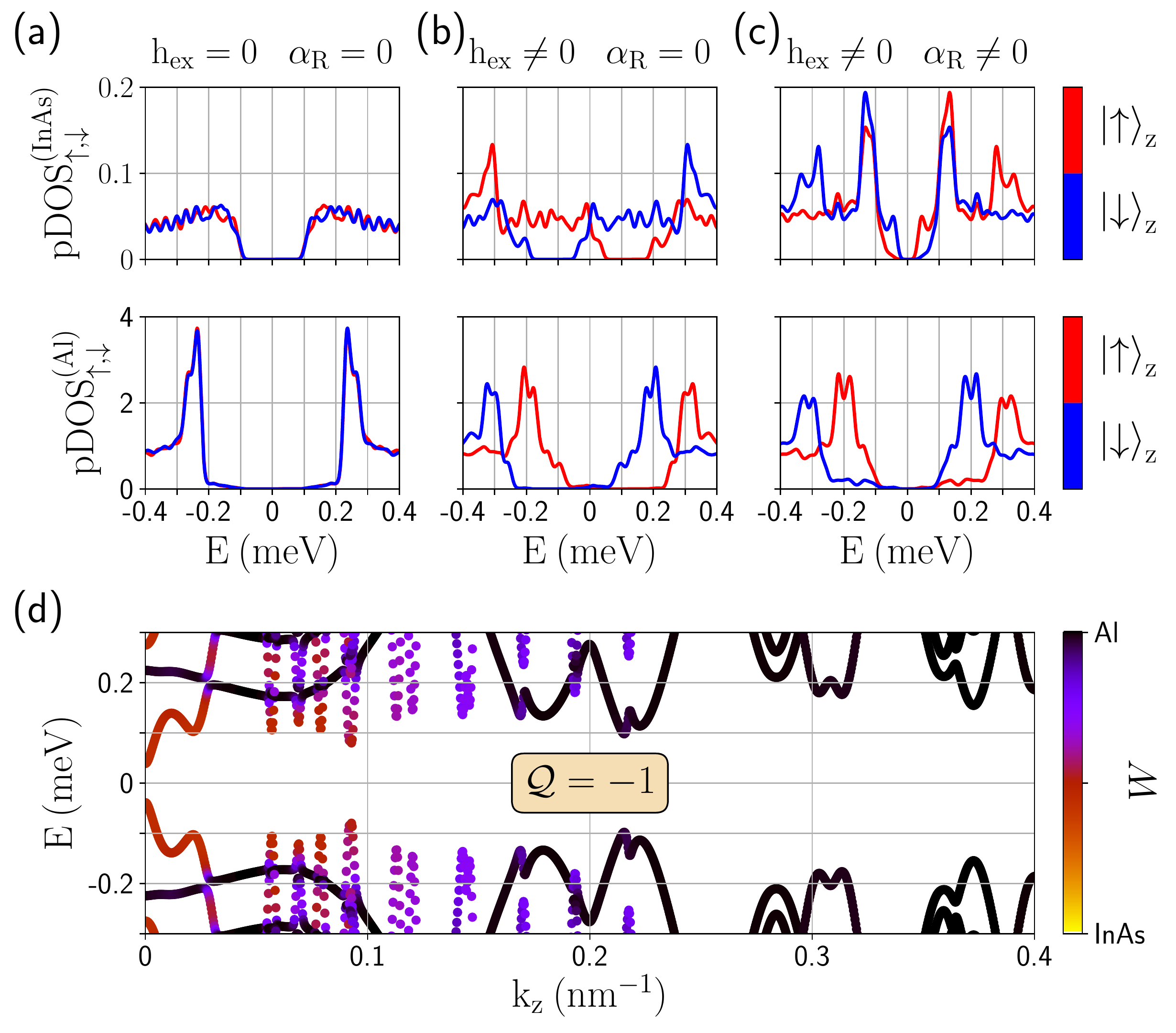}
\caption{\textbf{Full model results.} (a-c) Spin-resolved {partial density of states (pDOS) for the overlapping device integrated over the InAs wire volume (top row) and the Al layer volume} ({bottom} row) when (a) the exchange field $h_{\rm{ex}}$ in the EuS layer and the Rashba SOC $\alpha_{\rm{R}}$ in the InAs wire are set to zero, (b) only the exchange field is turned on, and (c) both are present. Red and blue correspond to the pDOS for different spin orientations along the $z$ axis (wire's direction). (d) {Low-energy band structure versus $k_z$ for the hybrid-wire parameters in (c).} The colorbar represents the {relative} weight $W$ of a given state in the Al layer (black) and in the InAs wire (yellow). The wavefunction weight in the EuS layer is negligible since it is an insulating material. 
The $\mathbb{Z}_2$ topological invariant is $\mathcal{Q}=-1$, signaling a topological phase.
We take here $V_{\rm bg}=-0.95$ V and $V_{\rm sg}^{\rm (L,R)}=0$ V. {Other parameters can be found in Table I of the SM.}}
\label{Fig2}
\end{figure}

\begin{figure}
\includegraphics[width=1\columnwidth]{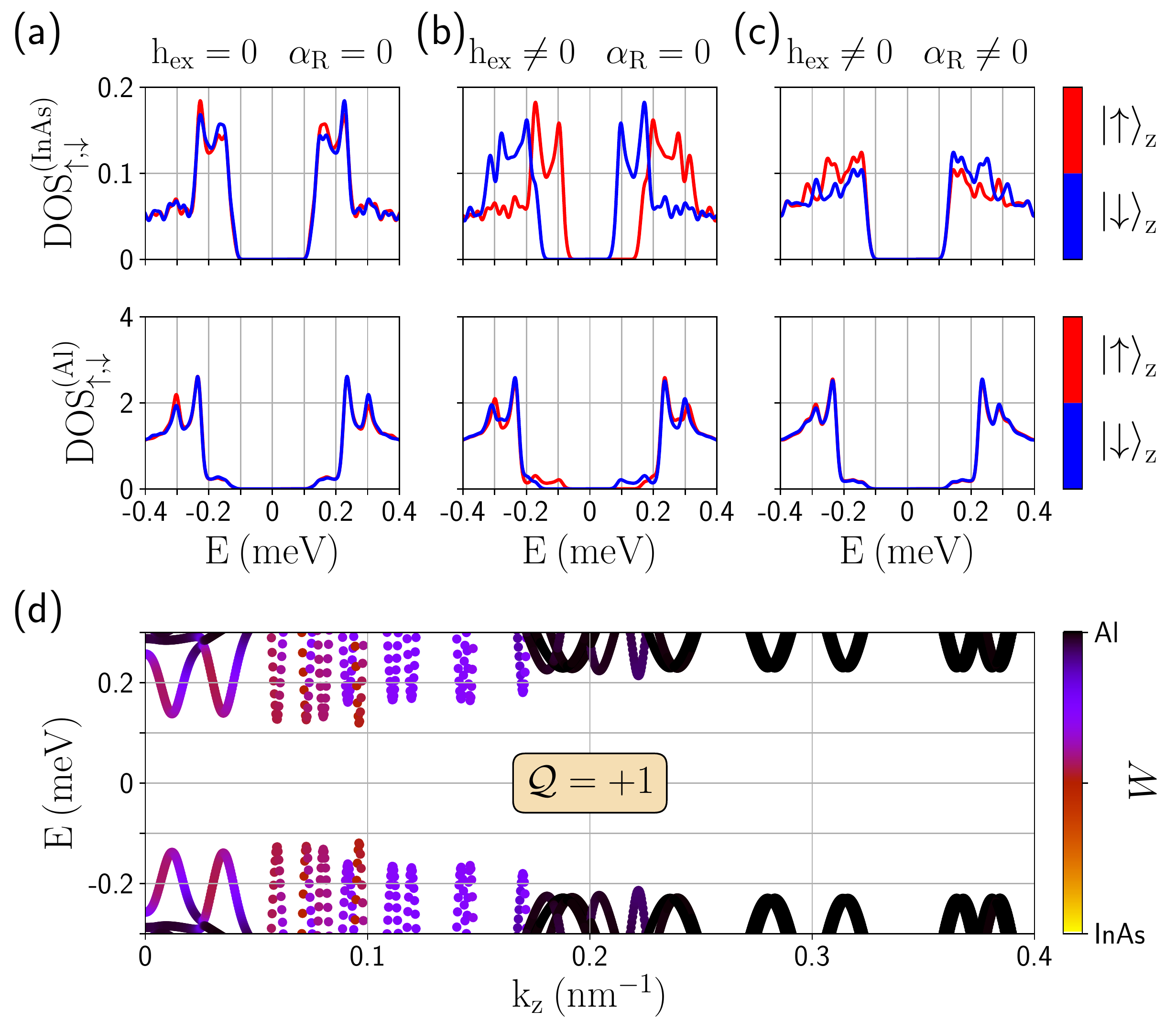}
\caption{Same as in Fig. \ref{Fig2} but for the non-overlapping device. {The $\mathbb{Z}_2$ topological invariant in (d) is now $\mathcal{Q}=1$, signaling a trivial phase.}}
\label{Fig3}
\end{figure}

{\it Full model results.---}
We first focus on the density of states (DOS) and {dispersion relation} of the overlapping geometry (see Fig. \ref{Fig2}) fixing the side-gate voltages to zero and the back-gate to $\sim -1$ V. In order to identify the separate effect of the magnetic and superconducting terms, we perform three different calculations: in the first one we switch off the exchange field in the EuS and the Rashba SOC in the InAs [Fig. \ref{Fig2}(a)]; then we switch on $h_{\rm{ex}}$ [Fig. \ref{Fig2}(b)] and finally we also connect $\alpha_{\rm{R}}$ [Fig. \ref{Fig2}(c)]. 

In the top panel of Fig. \ref{Fig2}(a) we show the partial DOS integrated over the InAs volume. It exhibits a well-defined induced superconducting gap, although halved with respect to the $\sim 0.2$ meV gap observed in the DOS integrated over the Al shell volume, Fig. \ref{Fig2}(a) bottom panel. This is in accordance with what one expects from a conventional superconducting proximity effect \cite{Chang:Nnano15, Moor:NJP18, Winkler:PRB19}.

\begin{figure*}
\includegraphics[width=2\columnwidth]{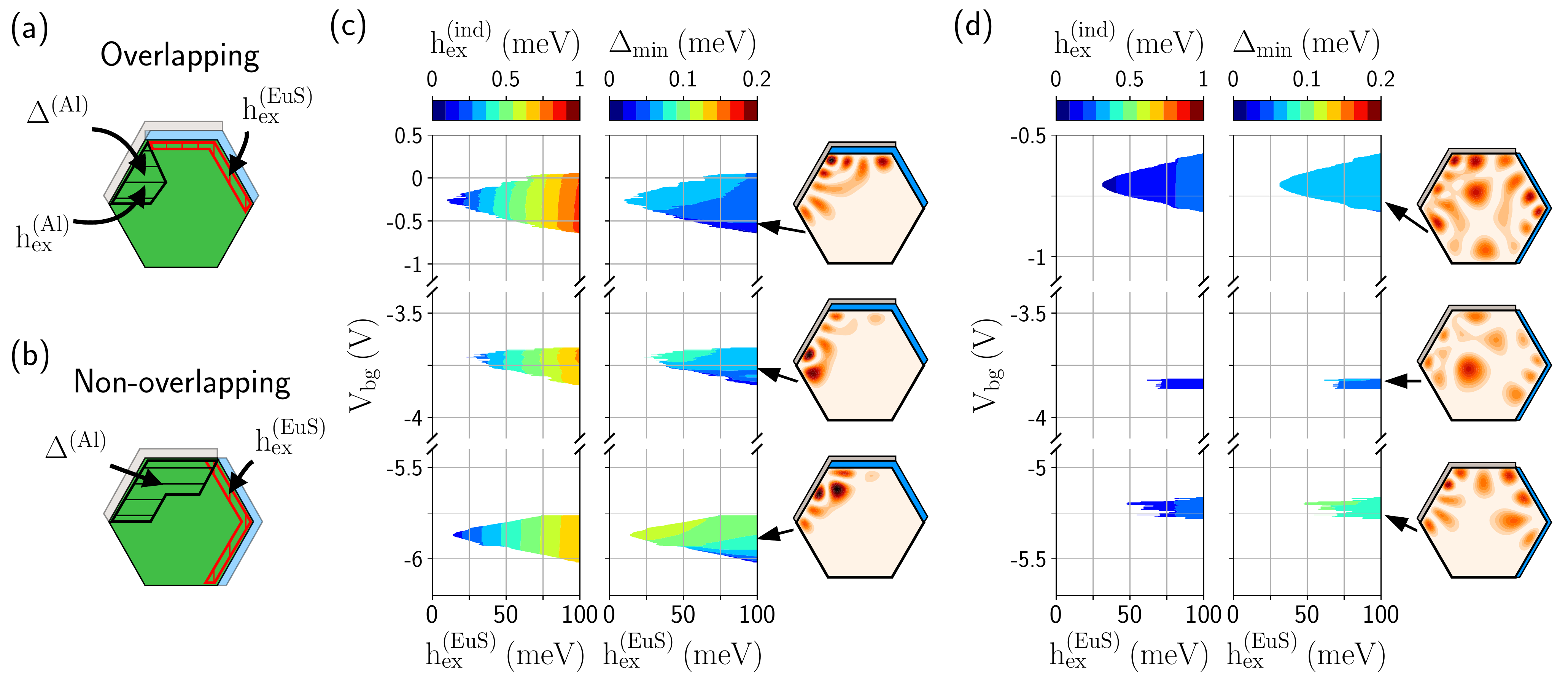}
\caption{\textbf{Simplified-model results.} {The Al and EuS layers are integrated out and their respective effective induced pairing amplitude and exchange field on the InAs wire are included within the streaked regions shown in the sketches of (a) the overlapping device and (b) the non-overlapping one. We take $\Delta^{\rm (Al)}=0.23$ meV and $h_{\rm ex}^{\rm (Al)}=0.06$ meV over a wide region of 45 nm near the Al interface. $h_{\rm ex}^{\rm (Al)}$ is only present in (a) where the Al and the EuS are in contact. We include $h_{\rm ex}^{\rm (EuS)}$ over a thin region of $1$ nm close to the EuS layer.} (c) Topological phase diagram of the overlapping device versus back-gate potential, $V_{\rm bg}$, and exchange field at the EuS-InAs interface, $h_{\rm ex}^{\rm (EuS)}$, for $V_{\rm sg}^{\rm (L,R)}=0$. {In the topological regions,} we show with colors the expectation value of the induced exchange field (left panel), the topological minigap (middle panel), and the wavefunction {profile} (right panel), all of them for the transverse subband closest to the Fermi energy. The  parameters for which the wavefunctions are plotted are indicated with arrows. (d) Same as (c) but for the non-overlapping device and fixing $V_{\rm sg}^{\rm (R)}=2$ V and  $V_{\rm sg}^{\rm (L)}=0$. The values of $V_{\rm sg}^{\rm (L,R)}$ in (c) and (d) are taken to maximize the topological regions in each case. Other parameters can be found in Table I of the SM \cite{Supplementary}. The extension of the topological phase is very much reduced with respect to (c) (almost negligible for some subbands) both in the $V_{\rm{bg}} $ and $h_{\rm ex}^{\rm (EuS)}$ axes. Moreover, in the regions where it is present, the topological gap is small. In contrast, large topological regions with stronger minigaps are found for the overlapping device. The reason is twofold: (i) the wavefunction can be pushed simultaneously close to the Al and EuS layers due to the electrostatics of the overlapped shells, which increases the superconducting and magnetic proximity effects; (ii) the induced exchange field feeds both from the direct and indirect contributions in this case. 
}
\label{Fig4}
\end{figure*}

When $h_{\rm{ex}}\neq 0$ (but $\alpha_{\rm{R}}=0$) we observe two main features in the spin-resolved partial DOS. First, an energy splitting of the superconducting coherence peak appears in the Al [Fig. \ref{Fig2}(b) bottom panel], which is of the order of $\sim 0.06$ meV, in agreement with recent theoretical and experimental results on Al/EuS junctions \cite{Hao:PRL91, Strambini:PRM17, Rouco:PRB19, Zhang:PRB20}. This agreement without any fine tuning of the parameters in our model is encouraging about its validity. Second, there is a complete closing of the induced gap in the InAs [Fig. \ref{Fig2}(b) top panel]. This points to an induced exchange field larger than $\sim 0.1$ meV, the induced gap in the semiconductor, and therefore, larger than in the Al layer. 
In contrast to previous proposals \cite{Liu:NanoLett20, Vaitiekenas:NatPhys20} our results suggest that in the current case topological superconductivity is achieved below the Chandrasekar-Clogston limit \cite{Chandrasekhar:AppPhysLet62, Clogston:PRL62} for the Al ($h_{\rm ex}^{\rm (Al)}<\Delta/\sqrt{2}$).


Finally, in Fig. \ref{Fig2}(c) top panel we observe that a gap is opened again in the presence of SOC. This sequence of gap closing and reopening at a high-symmetry $k_z$-point is a signature of a topological phase transition. The band structure shown in Fig. \ref{Fig2}(d) further illustrates the spatial distribution of the low-energy states in this last case (i.e., with $h_{\rm ex} \ne 0$ and $\alpha_{\rm R}\ne 0$). The weight $W$ of each state {in the different materials is represented with colors, from a state completely located in the Al layer (black) to one completely located in the InAs wire (yellow). The lowest-energy states close to $k_z=0$ have significant weight both in the Al and in the InAs, as expected for a topological superconducting phase \cite{Winkler:PRB19}. We prove that the system in Figs. \ref{Fig2}(c) and \ref{Fig2}(d) is indeed in the topological regime by calculating the $\mathbb{Z}_2$ topological invariant $\mathcal{Q}$. For large Hamiltonian matrices, this can be achieved by computing the Chern number from the eigenvalues of the Wilson matrix, which only involves the lowest-energy eigenstates at the Brillouin zone borders \cite{Supplementary}. We find $\mathcal{Q}=-1$, which actually corresponds to the nontrivial case.

Strikingly, the same analysis for the non-overlapping geometry (Fig. \ref{Fig3}) reveals that the magnetic proximity effect in this case is not {strong} enough to close and reopen the superconducting gap in the wire. The reason for this behavior can be traced to the limited spin polarization induced in the nanowire for this geometry. Hence, there is no topological phase in this case, at least for this choice of gate voltages.

{\it Simplified model and phase diagram.---}
{We consider now the Hamiltonian of Eq. \eqref{hamiltonian}} restricted to the InAs wire, where we include {an effective} pairing amplitude $\Delta^{(\rm{Al})}=0.23$ meV and an exchange field $h_{\rm ex}^{(\rm{EuS})}=100$ meV on the cross-section regions closer to the Al and the EuS shells, respectively, as schematically depicted in Figs.~\ref{Fig4}(a) and (b). We also include a smaller exchange coupling $h_{\rm ex}^{(\rm{Al})}=0.06$ meV in the Al-proximitized region of the overlapping device. The magnitude of these parameters and the extension of the corresponding regions are extracted by adjusting to the behavior of the full model results, as shown in the SM \cite{Supplementary}. 


In Fig. \ref{Fig4}(c) we present the topological phase diagram of the overlapping device as a function of the back-gate voltage and the exchange field of the EuS. Notice that $h_{\rm ex}^{\rm (EuS)}$ should be $100$ meV according to our full model. However, departures from the idealized model of Eq. \eqref{hamiltonian} might reduce the value of the induced magnetic exchange. For instance, the mismatch between the minima of the InAs conduction band (at the $\Gamma$ point) and the EuS one (at the $\mathrm{X}$ point \cite{Nirpendra:PBCM07}) could suppress their hybridization depending on material growth directions or other details (see the SM \cite{Supplementary}), leading to a smaller $h_{\rm ex}^{\rm (EuS)}$ value. Thus we allow this parameter to vary between $0$ and $100$ meV to evaluate the robustness of our results with respect to this value. 
With colors, in the left panel of Fig. \ref{Fig4}(c) we show the induced exchange coupling, $h_{\rm ex}^{\rm (ind)}=\left<h_{\rm ex}(\vec{r})\sigma_z\right>$ \footnote{Note that the induced exchange coupling in the wire, $h_{\rm ex}^{\rm (ind)}$, is much smaller (approximately one per cent \cite{Supplementary}) than the EuS parent Zeeman field, $h_{\rm ex}^{(\rm{EuS})}$.}, and in the middle panel the induced minigap, $\Delta_{\rm min}=\left|E(k_z=k_{\rm F})\right|$, for the energy state closest to the Fermi energy in both cases. In these plots, white means trivial (i.e., $\mathcal{Q}=1$), while the colored regions correspond to the topological phase. There are several topological regions against $V_{\rm bg}$ corresponding to different transverse subbands. In those regions, the condition that $h_{\rm ex}^{\rm (ind)}$ is larger than the square root of the induced gap squared plus the chemical potential squared is fulfilled, as expected~\cite{Lutchyn:PRL10, Oreg:PRL10}.
To the right in Fig. \ref{Fig4}(c) we show the probability density of the transverse subband closer to the Fermi level at $k_z=0$ across the wire section for the parameters indicated with arrows. In the three cases exhibited, the wavefunction concentrates both around the left-upper facet covered by Al, and the top facet where the Al and EuS layers overlap. This is consistent with the requirement of maximizing simultaneously the superconducting and magnetic proximity effects. 

The phase diagram for the non-overlapping device is shown in Fig. \ref{Fig4}(d). The extension of the topological phase is very much reduced, almost negligible for some subbands.  It is interesting to observe that, for realistic gate potential values, the wavefunction needs to be very spread across the wire section in order to acquire the superconducting and magnetic correlations for the topological phase to develop. This in turn translates into narrow back-gate voltage ranges for which this is possible and small topological minigaps.

In the SM \cite{Supplementary} we further analyze the previous phase diagrams as a function of the right-side gate voltage, obtaining similar results.
{We also consider alternative geometries, nevertheless finding that the overlapping configuration of Fig. \ref{Fig1}(a) gives rise to more extended, robust and tunable topological regions for realistic parameters. 
In particular,} we find that the best way to optimize {the topological state} (i.e., increase its minigap) is by fixing a large negative back-gate potential and a small positive right-side gate potential. In doing so, the wavefunction is pushed towards the superconductor-ferromagnet corner of the wire and thus the superconducting and magnetic proximity effects are maximized.


{\it Conclusions.---} From calculations of the DOS, band structure, topological invariant and the phase diagram, we conclude that the hybrid InAs/Al/EuS nanowires studied in Ref. \onlinecite{Vaitiekenas:NatPhys20}  can exhibit topological superconductivity under certain geometrical and gating conditions. 
For a topological phase to exist, the nanowire wavefunction must acquire both superconducting and magnetic correlations such that the induced exchange field exceeds the induced pairing. Since the proximity effects occur only in wire cross-section regions close to the Al and EuS layers, the wavefunction needs to be pushed simultaneously close to both materials by means of nearby gates. Our numerical simulations demonstrate that this is electrostatically favorable in device geometries where the Al and EuS shells overlap over some wire facet. This configuration is further advantageous in that, apart from a direct magnetization from the EuS layer in contact to the wire, there is an indirect one through the Al layer, which favors reaching the topological condition. While our model includes the effect of disorder at the Al layer surface and at the EuS/Al interface, we have not considered other sources of disorder, like ,e.g., the presence of magnetic domains. However, these domains can be aligned by the application of a small field that is then switched to zero, as done in Ref. \cite{Vaitiekenas:NatPhys20}. A subsequent study which considers a fully diffusive Al layer \cite{Antipov} reaches similar conclusions as our work (although disorder increases the induced exchange field required to achieve a topological phase).


Finally, as a side outcome, our microscopic analysis demonstrates the tunability of the magnetic and superconducting induced couplings in the nanowire. This opens up the possibility of engineering the material geometries, their disposition, and the electrostatic environment to enter and abandon the topological regime at will and, thus, the appearance of Majorana modes. These ideas can be applied to other materials and experimental arrangements, which surely will favor more experiments in the field, as well as in other fields such as superconducting spintronics.

{\it Acknowledgments.---}
We would like to thank S. Vaitiek\.enas and C. M. Marcus for illuminating discussions on their experimental results. We also thank M. Alvarado and O. Lesser for valuable help with the topological invariant calculation. A.L.Y. and E.P. acknowledge support from the Spanish MICINN through Grants No. FIS2016-80434-P (AEI/FEDER, EU) and No. FIS2017-84860-R, by EU through Grant No. 828948 (AndQC), and through the “Mar\'ia de Maeztu” Programme for Units of Excellence in R\&D (Grant No. MDM-2014-0377). Y.O. acknowledges partial support through the ERC under the European Union’s Horizon 2020 research and innovation programme (Grant Agreement LEGOTOP No. 788715), the ISF Quantum Science and Technology (2074/19), the BSF and NSF (2018643), and the CRC/Transregio 183.

\bibliography{EuS}


\appendix
\section*{Supplemental Material}

\section*{Numerical implementation}
\label{App:A}
In this section we describe the specific details of the two geometries (overlapping and non-overlapping) considered in the main text, as well as their characteristic material parameters. In addition, we explain how we perform the numerical
calculations for their electronic structure including the effect of the electrostatic potential.

\subsection*{1. Geometry}
A sketch of the two devices analyzed in this work is shown in Fig. 1 of the main text. In both cases a  hexagonal cross-section InAs nanowire (in green) of radius $60$ nm is partially covered by two different thin layers. Both are grown epitaxially over certain facets of the wire. The highly transparent epitaxial interfaces improve the hybridization between the nanowire and the layers.
One of these layers is a 6 nm-thick shell made of superconducting Al (light grey), whose outer surface (2 nm-thick) is oxidized. The other layer is made of EuS (blue), a well-known ferromagnetic insulator, and its width is 8 nm in the overlapping device and 5 nm in the non-overlapping one. Despite the difference between these two values, we find that the precise thickness of the EuS, for the ferromagnetic exchange coupling mechanism studied here, plays a negligible role because it acts as a mere (magnetic) insulator. We note that, in the overlapping device, the interface between the Al and the EuS is also grown epitaxially in the experiments but the interface is not completely flat. This roughness is characterized by a corrugation width of $\sim 1$ nm.

This hybrid system is surrounded by three gates (dark grey), one at the bottom and two at the left and right sides, which allow to tune the wire doping level and the charge distribution inside the heterostructure simultaneously. The side gates are placed symmetrically from the center of the wire at $30$ nm distance from the outer corners of the wire, and their height is half the wire's width. On the other hand, the wire is placed on top of a SiO$_2$ dielectric substrate (violet) with a thickness of $200$ nm. This substrate fixes the distance between the back gate and the bottom of the wire. 

There are other dielectrics in the devices that are necessary to avoid the unwanted oxidation of the EuS layer. They are a 8 nm-thick layer of HfO$_2$ (orange) surrounding the whole heterostructure in the overlpaping device, and a $10$ nm-thick layer of AlO$_2$ (yellow) covering the EuS layer in the non-overlapping one. We include them in our electrostatic simulations to obtain a better qualitative agreement with the experiment of Ref. \onlinecite{Vaitiekenas:NatPhys20} although they play a minor role in our results. Finally, we assume that the rest of the environment (white) is vacuum.
We summarize all these geometrical parameters in Table \ref{Table:parameters}.

\begin{table}[t]
\caption{Parameters used for the realistic calculations of the main text. All widths are extracted from Ref. \onlinecite{Vaitiekenas:NatPhys20}. Temperature is fixed to $100$ mK in all our simulations.}

\renewcommand{\arraystretch}{1.15}
\begin{tabularx}{\columnwidth} {X >{\centering}X >{\centering}X >{\centering\arraybackslash}X}
\hline \hline
\textbf{Material} & \textbf{Parameter} & \textbf{Value} & \textbf{Refs.} \\ 
\hline
InAs & width & 120 nm & - \\
    & height & 104 nm &     \\ \cline{2-4}
    & $m_{\rm eff}$ & 0.023$m_0$  & \onlinecite{Vurgaftman:JAP01} \\
    & $E_{\rm F}$  & 0  & \onlinecite{Liu:AppMat20}  \\
    & $h_{\rm ex}$ & 0  & - \\
    & $\Delta$ & 0  & - \\
    & $\alpha$ & \multicolumn{2}{c}{(see text for details)} \\ \cline{2-4}
    & $\epsilon_{\rm InAs}$ & 15.5$\epsilon_{0}$ & \onlinecite{Levinshtein:00}  \\
    & $\rho_{\rm surf}$ & $2\cdot 10^{-3}\mathrm{\left(\frac{e}{nm^{3}}\right)}$ & \onlinecite{Thelander:Nano10}, \onlinecite{Reiner:PRX20} \\ \hline
Al & width & 6 nm & - \\
    & oxidation width & 2 nm & \\ \cline{2-4}
    & $m_{\rm eff}$ & $m_0$  & \onlinecite{Segall:PR61} \\
    & $E_{\rm F}$  & -11.7 eV  &  \onlinecite{AschroftMermin:76} \\
    & $h_{\rm ex}$ & 0  & - \\
    & $\Delta$ & 0.23 meV  & \onlinecite{Chang:Nnano15} \\
    & $\alpha$ & 0  & \onlinecite{Chiang:EPL17} \\ \cline{2-4}
    & $\epsilon_{\rm Al}$ & $\infty$ & - \\    
    & $V_{\rm Al}$ & 0.2 eV & \onlinecite{Reiner:PRX20}, \onlinecite{Schuwalow:arxiv19}, \onlinecite{Wimmer}  \\ \hline
EuS & width\footnote{Note that in the experiments, the width of the EuS layer is $8$ nm in the overlapping device while it is $5$ nm in the non-overlapping one.} & 8/5 nm & - \\
    & corrugation width & 1nm & \onlinecite{Liu:NanoLett20} \\ \cline{2-4}
    & $m_{\rm eff}$ & 0.3$m_0$  & \onlinecite{Xavier:PLA67} \\
    & $E_{\rm F}$  & 0.8 eV  & \onlinecite{Alphenaar:09}, \onlinecite{Liu:AppMat20}  \\
    & $h_{\rm ex}$ & 0.1 eV  & \onlinecite{Mauger:PhysRep86}, \onlinecite{Alphenaar:09} \\
    & $\Delta$ & 0  & - \\
    & $\alpha$\footnote{We have found no explicit information about the SO coupling in EuS materials. We therefore assume that it is negligibly small.} & 0  & - \\ \cline{2-4}
    & $\epsilon_{\rm EuS}$ & 10$\epsilon_{0}$ & \onlinecite{Axe:JPCS69}  \\ \hline
Dielectrics & SiO$_2$ width & 200 nm & - \\
    & HfO$_2$ width & 80 nm & \\
    & Al$_2$O$_3$ width & 8 nm & \\ \cline{2-4}
    & $\epsilon_{\rm SiO_2}$ & 3.9$\epsilon_{0}$ & \onlinecite{Robertson:EPJAP04} \\
    & $\epsilon_{\rm HfO_2}$ & 25$\epsilon_{0}$ & \onlinecite{Levinshtein:00} \\
    & $\epsilon_{\rm Al_2O_3}$ & 10$\epsilon_{0}$ & \onlinecite{Levinshtein:00} \\
    & $\epsilon_{\rm vacuum}$ & $\epsilon_{0}$ & -  \\
\hline \hline
\end{tabularx}

\label{Table:parameters}
\end{table}

\subsection*{2. Model Hamiltonian}
The basic arguments leading to our continuum model Hamiltonian have been presented in the main text. As explained there, we focus on the InAs, Al and EuS conduction bands and neglect the InAs and EuS valence bands. Notice that the EuS conduction band has its minimum at the $\mathrm{X}$-point \cite{Nirpendra:PBCM07} rather than at $\Gamma$, as InAs and Al do. When electrons tunnel from one material to the other, for instance from InAs to EuS, this could lead to momentum mismatch depending on the growth direction between both materials and thus to a decreased coupling. Moreover, the EuS could exhibit grains (or other forms of disorder), details which are beyond the qualitative analysis in this work. For this reason, we consider an effective continuum model for the different materials where the EuS conduction band minimum is also taken at the $\Gamma$-point. We note that other sources of momentum mismatch coming from different effective masses, finite size materials, certain sources of disorder explained below, etc., are taken into account. This model, which we call \textit{full model} in the main text, gives an optimistic hybridization between EuS and InAs. Since the boundary between the EuS and Al is disordered (see below), our model essentially describes correctly this hybridization as momentum conservation along that interface breaks under such disorder.


Under the assumptions presented above, we arrive to the following Bogoliubov-de Gennes Hamiltonian to describe the heterostructure (Eq. (1) in the main text)
\begin{eqnarray}
H &=& \left[ \vec{k}^T  \frac{\hbar^2}{2m_{\rm eff}(\vec{r})} \vec{k} + e\phi(\vec{r}) - E_{\rm F}(\vec{r}) \right]\sigma_0\tau_z \nonumber\\ 
&+& h_{\rm ex}(\vec{r})\sigma_z\tau_z + \Delta(\vec{r})\sigma_y\tau_y  \nonumber\\ 
&+& \frac{1}{2} \left[\vec{\alpha}(\vec{r})\cdot \left(\vec{\sigma}\times\vec{k}\right) + \left(\vec{\sigma}\times\vec{k}\right)\cdot \vec{\alpha}(\vec{r}) \right]\tau_z,
\label{Eq:Hamiltonian}
\end{eqnarray}
where $\vec{k}=(k_x,k_y,k_z)$ is the momentum operator vector, $\vec{r}=(x,y)$ is the radial spatial operator vector (for a translational-invariant wire along $z$), and $\sigma$ and $\tau$ are the Pauli matrices in spin and Nambu space, respectively. The first term in Eq. \eqref{Eq:Hamiltonian} is the kinetic energy, where $m_{\rm eff}(\vec{r})$ is the effective mass that takes different constant values inside the different parts of the heterostructure. The function $\phi(\vec{r})$ is the spatial-dependent electrostatic potential across the section of the hybrid nanowire. In the following subsection we explain in detail how we compute it. The quantity $E_{\rm F}(\vec{r})$ is the energy difference at $\phi=0$ between the band bottom of the conduction band (in each region) and the Fermi level of the entire heterostructure, which we fix to zero in this work. To simulate the disordered outer surface of the Al (the oxidized region) and the irregular EuS/Al interface, we assume that $E_F(\vec{r})$ in these disordered regions is characterized by a random Gaussian noise with a standard deviation of 10\% of the Fermi energy in the material. This accidental disorder breaks the parallel momentum conservation leading to an enhanced hybridization between the wire's and the layers' electronic states \cite{Winkler:PRB19}. The operators $h_{\rm ex}(\vec{r})$ and $\Delta(\vec{r})$ are the exchange field and the superconducting pairing amplitude, which take non-zero values only in the EuS and Al layers, respectively. Finally, the last term in Eq. \eqref{Eq:Hamiltonian} is the spin-orbit (SO) interaction, where $\alpha$ is the so-called SO coupling. In the last subsection we discuss separately how we describe it for the whole heterostructure. The different values for the parameters that we use in our simulations, together with the references from where they are extracted, are shown in Table \ref{Table:parameters}.

We perform a numerical diagonalization of the Hamiltonian \eqref{Eq:Hamiltonian} to obtain the eigenspectrum. To this end, we discretize this continuum model into a tight-binding Hamiltonian. We perform the discretization only in the $x$ and $y$ directions (the section of the wire), and we assume that in the other direction (the wire's direction $z$) the nanowire is translationally invariant. In the directions $x$ and $y$ the momentum must be substituted by their corresponding partial derivatives, i.e., $k_x\rightarrow -i\partial_x$ and $k_y\rightarrow -i\partial_y$, and then the partial derivatives can be discretized using the central difference method, e.g.
\begin{equation}
k_x\rightarrow -i\frac{\partial}{\partial x}\rightarrow -i\frac{c_{i+1}^{\dagger}c_i-c_{i-1}^{\dagger}c_i}{2h},
\end{equation}
where $h$ is the mesh spacing. We choose $h$ such that the grid can accommodate the Fermi wavelength of the eigenstates. Otherwise, a poor spacing would lead to spurious solutions. As Al is a metal with a large Fermi energy, it is characterized by a large Fermi momentum, which implies that a small mesh spacing must be used. We find that $h \sim 0.1$ nm is a reasonable mesh size in our simulations. Once the Hamiltonian is transformed into a tight-binding one, we diagonalize it numerically (for each $k_z$ and $\phi(\vec{r})$ values) using the standard ARPACK tools provided by the package \textit{Scipy} of Python. This gives as a result the energy subbands $E^{(j)}(k_z)$ for the different transverse subbands index $j$, and their corresponding wavefunctions $\Psi^{(j)}(x,y,k_z)$, which are actually four-component spinors. We perform this diagonalization using the Python package \textit{MajoranaNanowires} \cite{MajoranaNanowiresQSP_v1}.

As the three materials that comprise the hybrid wire are coupled, the energy subbands correspond to wavefunctions that spread all across the heterostructure. To characterize the weight $W_{R}$ of the wavefunction on a certain material $R$, we project the probability density of any state $j$ onto this region,
\begin{equation}
W_{R}^{(j)}(k_z)=\int_{\vec{r}\in R} dx \; dy \left| \Psi^{(j)}(x,y,k_z)\right|^2.
\end{equation}
In close relation to this quantity, we also compute the spin-polarized partial density of states (pDOS), i.e., the DOS integrated over different regions $R$, using the following expression
\begin{eqnarray}
\mathrm{pDOS_{\uparrow,\downarrow}^{(R)}}(E)=\sum_{j}\int_{\vec{r}\in R} dx \; dy \int_{k_z} dk_z \nonumber \\ \cdot \delta(E-E^{(j)}(k_z)) \left| \Psi^{(j)}_{\uparrow,\downarrow}(x,y,k_z)\right|^2. \: \:
\end{eqnarray}
If $R$ includes the three materials in this equation, then the resulting expression is the total spin-polarized $\mathrm{DOS}_{\uparrow,\downarrow}$ in the device. Then, one can compute the total DOS by summing both spin directions, $\mathrm{DOS_T=DOS_{\uparrow}+DOS_{\downarrow}}$.

Lastly, to characterize the $D$-class topological phase \cite{Atland:PRB97, Schnyder:PRB08} (to which the Majorana nanowires belong), we compute the Chern number
\begin{equation}
C=\frac{1}{2\pi i}\int_{R_{k_z}}dk_z \left< \nabla_{k_z} \Phi_{k_z} \left| \times \right| \nabla_{k_z} \Phi_{k_z} \right>,
\end{equation}
where $R_{k_z}$ denotes the first Brillouin zone ($k_z \in (-\pi,\pi]$ in our system), and $\Phi_{k_z}$ is the ground state of the system, i.e., the Slater determinant of all the $j$-th eigenstates whose energies $E^{(j)}$ are below the Fermi energy. To easily compute numerically the Chern number we perform some transformations following the ideas of Refs. \onlinecite{Loring:IOP10, Zhang:IOP13, Lesser:PRR20}. First, we apply the Stokes theorem
\begin{equation}
C=\frac{1}{2\pi i}\oint_{\partial R_{k_z}}dk_z \left< \Phi_{k_z} | \nabla_{k_z} \Phi_{k_z} \right>,
\end{equation}
in such a way that the (line) integral is now performed through a closed contour around the reciprocal unit cell. Then, we discretize the momentum space $k_z\rightarrow k_l$, what allows us to replace the derivatives by finite differences and the integral by a summation over the reciprocal space points
\begin{eqnarray}
C=\frac{1}{2\pi i}\sum_l \frac{\left< \Phi_{k_l} | \Phi_{k_{l+1}} \right> - \left< \Phi_{k_l} | \Phi_{k_{l-1}} \right>}{2(k_{l+1}-k_l)}=\nonumber \\
\frac{1}{2\pi} \mathrm{Arg}\left\{\prod_l \left< \Phi_{k_l} | \Phi_{k_{l+1}} \right>\right\},
\end{eqnarray}
where $\mathrm{Arg}\left\{z\right\}\equiv \Im\left\{\log \left(\frac{z}{|z|}\right) \right\}$ is the complex argument function of a complex number $z$. The product of Slater determinants $\left< \Phi_{k_l} | \Phi_{k_{l+1}} \right>$ in the above equation can be rewritten as a determinant of a matrix $\tilde{C}_{k_l,k_{l+1}}$, whose matrix element $(i,j)$ can be directly written in terms of the eigenfunctions of the Hamiltonian,
\begin{equation}
\tilde{C}_{k_l,k_{l+1}}^{i,j}=\left< \Psi^{(i)}(k_l) | \Psi^{(j)}(k_{l+1}) \right>.
\label{Eq:C_def}
\end{equation}
From this, the Chern number is simply given by
\begin{eqnarray}
C= \frac{1}{2\pi} \mathrm{Arg}\left\{ \det\left(\prod_l  \tilde{C}_{k_l,k_{l+1}} \right)\right\} \equiv \nonumber \\
 \frac{1}{2\pi} \mathrm{Arg}\left\{ \det(\mathcal{W})\right\},
\end{eqnarray}
where $\mathcal{W}$ is the so-called Wilson matrix in the literature \cite{Yu:PRB11, Alexandradinata:PRX16, Bradlyn:PRB19}. Finally, we use that the determinant of any square matrix can be written as the product of its eigenvalues to rewrite the expression as
\begin{equation}
C=\frac{1}{2\pi} \sum_l \mathrm{Arg}\left\{ \lambda_l\right\},
\label{Eq:Chern_number}
\end{equation}
where $\lambda_l$ are all the eigenvalues of the Wilson matrix $\mathcal{W}$. In practice, it is enough to include only the high-symmetry $k$-points in the product inside the Wilson matrix. In our case, these points are $k_z=\left\{-\pi,0,\pi\right\}$, which gives as a result the Wilson matrix
\begin{equation}
\mathcal{W}=\tilde{C}_{-\pi,0}\tilde{C}_{0,\pi}\tilde{C}_{\pi,-\pi}.
\label{Eq:Wilson_matrix}
\end{equation}
Additionally, notice that in principle the matrix $\tilde{C}_{k_l,k_{l+1}}$ involves all the eigenstates of the Hamiltonian (below the Fermi level) at two different $k_z$ points, $k_l$ and $k_{l+1}$, as shown in Eq. \eqref{Eq:C_def}. But since only the non-trivial topological eigenstates provide a non-zero contribution to the Chern number, and these states can only emerge close to the Fermi level in the studied system, it is therefore enough to include the closest states to the Fermi level in the Wilson matrix. This assumption allows to significantly alleviate the computational effort in our calculations.

To summarize this last part, the procedure that we follow in order to compute the topological invariant is the following: We first calculate the eigenstates $\Psi^{(j)}(k_z)$ of the Hamiltonian at $k_z=\left\{0,\pi,-\pi\right\}$. We need only those $j$-th states that are below and close to the Fermi energy. We find that $200$ eigenstates are enough in our system to properly characterize the topological phase. We then compute the coupling matrices $\tilde{C}_{k_l,k_{l+1}}$ of Eq. \eqref{Eq:C_def} and, from them, the Wilson matrix $\mathcal{W}$ using Eq. \eqref{Eq:Wilson_matrix}. We calculate its eigenvalues $\lambda_i$, and from them we compute the Chern number using Eq. \eqref{Eq:Chern_number}, which is a positive integer. We finally identify the $\mathbb{Z}_2$ topological invariant
\begin{equation}
\mathcal{Q}=(-1)^{C},
\end{equation}
which is positive (+1) for the trivial topological phase and negative (-1) for the non-trivial one.

\subsection*{3. Electrostatic potential}
We obtain the electrostatic potential $\phi(\vec{r})$ by solving the Poisson equation across the section of the wire,
\begin{equation}
\vec{\nabla}\left(\epsilon(\vec{r})\vec{\nabla}\phi(\vec{r})\right)=-\rho(\vec{r}),
\label{Eq:Poisson}
\end{equation}
where $\epsilon(\vec{r})$ is the dielectric permitivity, and $\rho(\vec{r})$ is the charge inside the heterostructure. On the one hand, $\epsilon(\vec{r})$ is characterized by a different constant value inside each material (see Table \ref{Table:parameters}), leading to abrupt jumps at the interfaces. Note that the Poisson equation must be solved on the entire system, including the dielectric environment. On the other hand, the charge inside the hybrid nanowire is given by,
\begin{equation}
\rho(\vec{r})=\rho_{\mathrm{surf}}(\vec{r})+\rho_{\mathrm{mobile}}(\vec{r}).
\label{Eq:total_charge}
\end{equation}
The term $\rho_{\mathrm{surf}}$ models the surface charge that is typically present in the uncovered nanowire facets \cite{Thelander:Nano10, Reiner:PRX20}. We model it as a \textit{positive} charge-density layer of thickness 1 nm in every nanowire facet not covered by the Al \cite{Winkler:PRB19, Escribano:PRR20}. This leads to a band bending at these facets and, thus, to an accumulation of charge in the wire in those regions [see Fig. 1(c)]. In principle, the wire facets covered by EuS should not display this type of charge accumulation layer because the EuS is grown epitaxially. However, recent experiments \cite{Liu:AppMat20} showed that there is a band bending at the EuS/InAs interface of the same magnitude as the one between InAs and air. Hence, we model both phenomena with the same $\rho_{\mathrm{surf}}$. We find that $\rho_{\mathrm{surf}}=2\cdot 10^{-3}\mathrm{\left(\frac{e}{nm^{3}}\right)}$ creates a band bending of 0.1 eV, which is the value reported in experiments for InAs/vacuum \cite{Thelander:Nano10, Reiner:PRX20} and InAs/EuS \cite{Liu:AppMat20} interfaces.

\begin{figure}
\includegraphics[width=1\columnwidth]{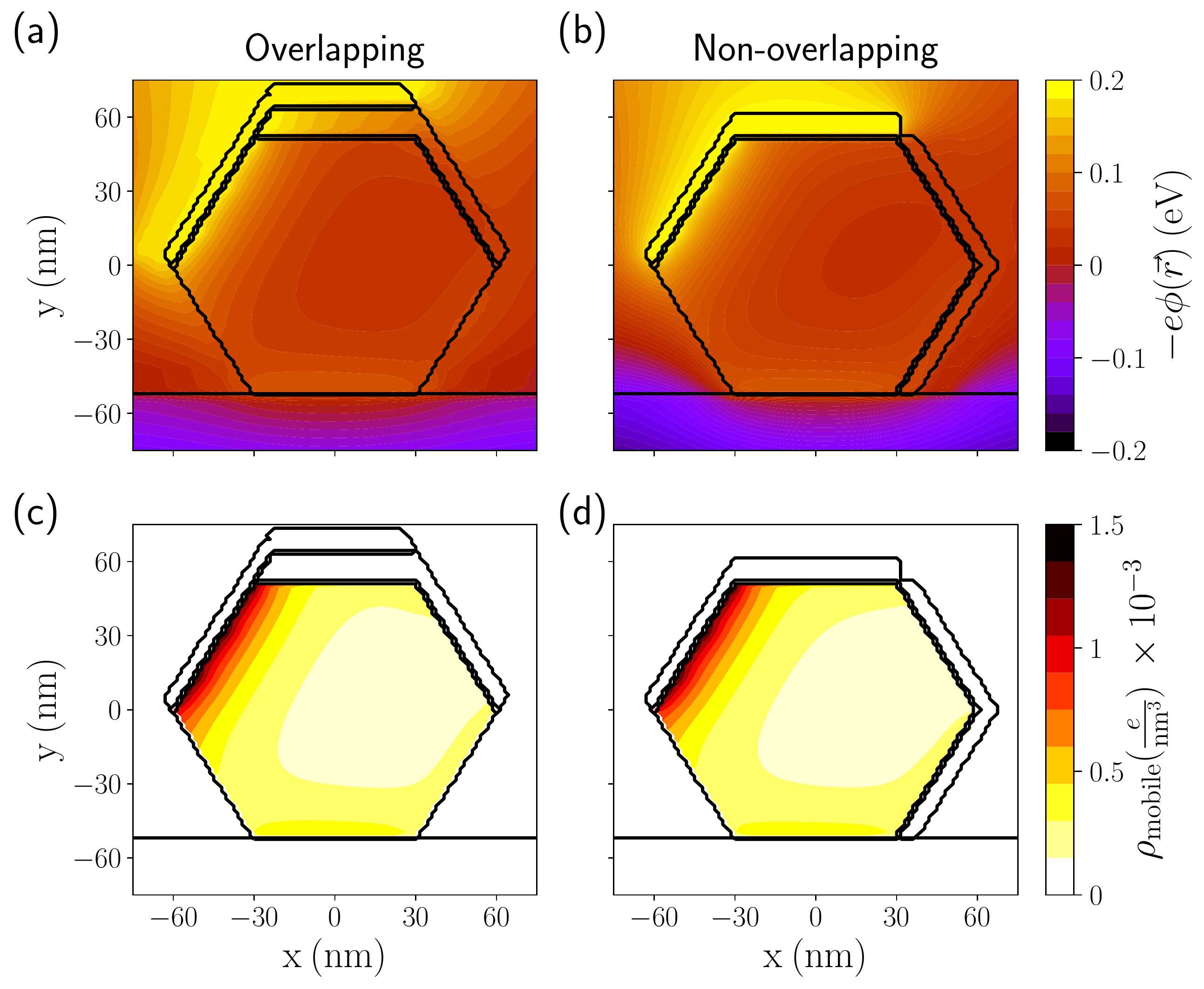}
\caption{Electrostatic potential across the heterostructure section in the overlapping device (a) and in the non-overlapping one (b). (c,d) Mobile charge density in the wire computed within the Thomas-Fermi approximation using the upper electrostatic potentials. Same parameters as in Figs. 2 and 3 of the main text, respectively.}
\label{FigA3}
\end{figure}

The second term in Eq. \eqref{Eq:total_charge} corresponds to the mobile charges inside the hybrid structure. In principle, they should be computed from the solutions of the Schr\"odinger equation using the Hamiltonian of Eq. \eqref{Eq:Hamiltonian}. Since this equation depends in turn on the electrostatic potential given by the Poisson equation of Eq. \eqref{Eq:Poisson}, the Schr\"odinger-Poisson equations have to be solved together in a self-consistent way. To reduce the computational effort, we use the Thomas-Fermi approximation. Within this approximation, one assumes that the charge density of the wire is given by the one of a non-interacting 3D electron-gas \footnote{We ignore the contribution coming from the valence bands as they could only play a role for $V_{\rm bg}\lesssim -4$ V in our simulations.}
\begin{eqnarray}
\rho_{\rm mobile}\simeq \rho_{\rm mobile}^{\rm (TF)}= \: \: \: \: \nonumber \\
 -\frac{e}{3\pi^2}\left[\frac{2m_{\rm eff}\left|e\phi(\vec{r})-E_F\right|f\left(-(e\phi(\vec{r})-E_F)\right)}{\hbar^2} \right]^{\frac{3}{2}}, \: \: \: \: \: \:
\label{Eq:Thomas-Fermi}
\end{eqnarray}
where $f(E)$ is the Fermi-Dirac distribution for a given energy $E$ and temperature $T$. This approximation has been used in previous works \cite{Mikkelsen:PRX18, Escribano:PRB19, Escribano:PRR20} on InAs nanowires and Al/InAs heterostructures, where it is demonstrated that it provides an excellent agreement with the results of the full Schr\"odinger-Poisson calculations. In this way, both equations are decoupled and, therefore, one only needs to diagonalize the Hamiltonian once for each potential $\phi(\vec{r})$ (and momentum $k_z$). 

Nonetheless, the Poisson equation still needs to be solved self-consistently [note that $\rho_{\rm mobile}^{\rm (TF)}$ depends on $\phi(\vec{r})$], although at a reduced computational cost. To carry out this self-consistency we use an Anderson iterative mixing,
\begin{equation}
\rho_{\rm mobile}^{(n)}=\beta\tilde{\rho}_{\rm mobile}^{(n)}+(1-\beta)\rho_{\rm mobile}^{(n-1)},
\end{equation}
where $n$ is a certain iteration in the self-consistent process, and $\beta$ is the so-called Anderson coefficient which is a constant in the range $[0,1]$. In the first step ($n=0$) we take $\rho^{(0)}=0$ and we compute the electrostatic potential using the Poisson equation. Then, at any other arbitrary step $n$, we compute the charge density $\tilde{\rho}_{\rm mobile}^{(n)}$ using Eq. \eqref{Eq:Thomas-Fermi} with the electrostatic potential found using the $\rho_{\rm mobile}^{(n-1)}$ of the previous iteration $n-1$. We finally find the charge density at the $n^{th}$ step (i.e., $\rho_{\rm mobile}^{(n)}$) using the Anderson mixing between these two charge densities shown in the above equation. This mixing between two different steps ensures the convergence to the solution. We keep this iterative procedure until the cumulative error between $\rho_{\rm mobile}^{(n)}$ and $\rho_{\rm mobile}^{(n-1)}$ is below $0.5\%$. To facilitate an efficient convergence, we choose as Anderson coefficient a self-adaptive one given by,
\begin{equation}
\beta=\beta^{(max)}\exp{\left( \frac{\max{\left(\left||\rho_{\rm mobile}^{(n)}|-|\rho_{\rm mobile}^{(n-1)}|\right|\right)}}{\max(|\rho_{\rm mobile}^{(n)}|,|\rho_{\rm mobile}^{(n-1)}|)} \right)},
\end{equation}
where $\beta^{(max)}$ is the maximum value that we allow the Anderson coefficient to take in order to ensure converge. We set this value to 0.1 in all our simulations.

To solve the Poisson equation, we also need to impose boundary conditions. We fix different potentials at the boundaries of the gates for the different simulations performed in this work. In addition to this, we fix a constant potential of $V_{\rm Al}=0.2$ eV at the boundaries with the Al shell. The electrostatic potential resulting from this last condition will create a band bending of the same magnitude at the InAs/Al interface [see Fig. 1(c)]. This band bending, reported in experiments \cite{Reiner:PRX20, Schuwalow:arxiv19, Wimmer}, has been previously described in other theoretical works \cite{Mikkelsen:PRX18, Antipov:PRX18, Escribano:BJN18, Winkler:PRB19, Escribano:PRB19, Woods:PRB20} on similar InAs/Al heterostructures.

To solve the Poisson equation with the above ingredients we use \textit{Fenics} \cite{Logg:10, Logg:12}, a finite element solver for Python. We use a mesh discretization of $1$ nm.

As we show in the main text, the precise geometries of the devices, the position of the Al and EuS shells on the InAs facets and the gates, play a key role in the appearance of a topological phase. To better illustrate this fact, we show in Fig. \ref{FigA3} the electrostatic potential in the overlapping (a) and the non-overlapping (b) devices. In these simulations, the back-gate voltage is set to a negative value, particularly $V_{\rm bg}=-1$ V. This is a typical situation in our simulations and in the experiments, as one needs to deplete the wire in order to populate it with just a few bands. Hence, the electrostatic potential is negative at the bottom of the wire, while it is positive and maximum close to the Al/InAs interface due to the Al/InAs band bending. This is translated into an accumulation of (mobile) charges $\rho_{\rm mobile}$ in this interface, quantity that we show in Fig. \ref{FigA3}(c) and (d) for the same devices. Despite the similarity between the electrostatic profiles of both devices, we show in the main text that they give rise to a very different energy spectrum. This can be understood by noticing the position of the the charge density with respect to the EuS. The better part of the charge density is located at the Al/InAs interface. In the overlapping device [Fig. \ref{FigA3}(c)], this charge is thus also close to the left part of the EuS layer (the top facet). On the contrary, the EuS layer in the non-overlapping device [Fig. \ref{FigA3}(d)] is far apart from the charge density. This apparently minor difference is what enhances the hybridization with the EuS in the overlapping device, and what suppresses the same hybridization in the non-overlapping one. We remark that this hybridization is necessary to induce a strong enough exchange field in the wire, and therefore, to create a topological phase.

\subsection*{4. Spin-orbit coupling}
\label{subsection-SOI}
The SO interaction included in the Hamiltonian of Eq. \eqref{Eq:Hamiltonian} arises whenever an spatial symmetry is broken. In this Hamiltonian, only linear terms with $k$ are included as they are known to be the dominant ones. The strength of the SO interaction is given by the SO coupling $\vec{\alpha}(\vec{r})$ and its value depends on the precise material. For Al, the SO coupling is negligible \cite{Chiang:EPL17}, while for the EuS we have found no information about its value in the literature. We thus set to zero the SO coupling of both materials in our simulations. By contrast, the SO coupling in InAs nanowires can take large values \cite{Campos:PRB18, Escribano:PRR20}. Because the topological protection of the Majorana nanowires depends strongly on the precise value of this interaction, a proper description is crucial to reach a good qualitative agreement with the experiments.

\begin{figure}
\includegraphics[width=1\columnwidth]{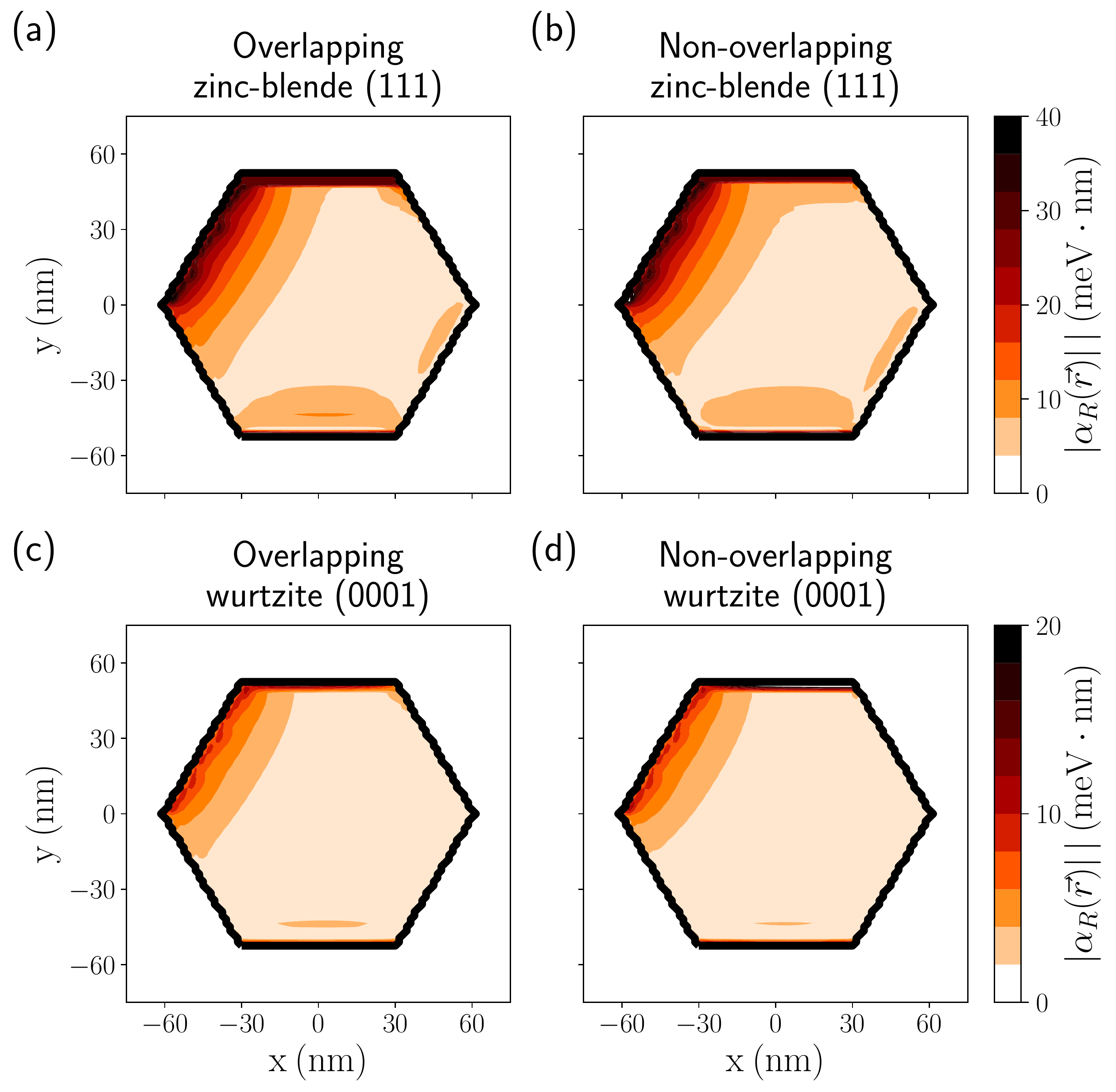}
\caption{Rashba SO coupling across the wire section for a (111) zinc-blende InAs nanowire (top) and a (0001) wurtzite InAs nanowire (bottom), in the overlapping device, (a) and (c), and in the non-overlapping one, (b) and (d). Same parameters as in Fig. \ref{FigA3}.}
\label{FigA4}
\end{figure}

The SO coupling has, in general, two components \cite{Winkler:03},
\begin{equation}
\vec{\alpha}(\vec{r})=\vec{\alpha}_{\rm D}+\vec{\alpha}_{\rm R}(\vec{r}),
\end{equation}
where $\alpha_{\rm D}$ is the Dresselhaus term and $\alpha_{\rm R}$ the Rashba one. The Dresselhaus SO coupling arises when the crystal unit cell itself is not symmetric, leading to a bulk inversion asymmetry. Hence, it is a spatial independent value that only depends on the material and its precise crystallographic structure. For InAs nanowires, it is known to be negligible in (111) zinc-blende crystals, while its value is roughly $\vec{\alpha}_{\rm D}=(0,0,30)$(meV$\cdot$nm) in (0001) wurtzite ones \cite{Gmitra:PRB16, Faria:PRB16}. On the other hand, the Rashba SO coupling emerges when the mesoscopic system is not symmetric in some direction, as it happens when there are interfaces or potential gates. It can be accurately described for III-V compound semiconductors using the following equation
\begin{eqnarray}
\vec{\alpha}_R (\vec{r})=\frac{eP_{\mathrm{fit}}^2}{3} \left[\frac{1}{\Delta_{\rm g}^2}-\frac{1}{(\Delta_{\rm{soff}}+\Delta_{\rm g})^2}\right] \vec{\nabla}\phi(\vec{r}),
\label{Eq:Rashba_SOC}
\end{eqnarray}
where $P_{\rm fit}$ is a parameter that depends on the material and its crystallographic structure, and $\Delta_{\rm g}$ and $\Delta_{\rm soff}$ are the valence band and split-off bands, respectively. We show in Table \ref{Table:parameters_SOC} their values. This equation has been derived in a previous work \cite{Escribano:PRR20}, finding that it provides excellent results in comparison to both more sophisticated theories and experimental measurements.

\begin{table}[t]
\caption{Parameters used for the SO coupling in the InAs nanowire, extracted from Ref. \onlinecite{Escribano:PRR20} and references therein. These parameters are valid for a 120 nm wide nanowire, except for $\vec{\alpha}_D$ that is valid for any width.}
\renewcommand{\arraystretch}{1.15}
\begin{tabularx}{\columnwidth} {X >{\centering}X >{\centering\arraybackslash}X}
\hline \hline
\textbf{Crystal} & \textbf{Parameter} & \textbf{Value} \\ 
\hline
(111) Zinc-blende & $\vec{\alpha}_{\rm D}$ & (0,0,0) \\
    & $P_{\rm fit}$ & 1300 meV$\cdot$nm \\ 
    & $\Delta_{\rm g}$ & 417 meV \\ 
    & $\Delta_{\rm soff}$ & 390 meV \\ \hline
(0001) Wurtzite & $\vec{\alpha}_{\rm D}$ & (0,0,30) meV$\cdot$nm\\
    & $P_{\rm fit}$ & 700 meV$\cdot$nm \\ 
    & $\Delta_{\rm g}$ & 467 meV \\ 
    & $\Delta_{\rm soff}$ & 325.7 meV \\
\hline \hline
\end{tabularx}

\label{Table:parameters_SOC}
\end{table}

We note that the Rashba SO coupling is a position-dependent function because it depends on the gradient of the electrostatic potential $\phi(\vec{r})$. To better illustrate this dependence, in Fig. \ref{FigA4} we show calculations of the SO coupling for the two devices studied in this work (first column and second column) and for (111) zinc-blende (first row) and (0001) wurtzite (second row) InAs nanowires. The Rashba SO coupling in zinc-blende nanowires (top row) turns out to be roughly twice than in wurtzite crystals (bottom row). This has been pointed out in other theoretical works \cite{Campos:PRB18, Escribano:PRR20} and it is intimately related to the symmetries of both crystals. But in both cases, the spatial profile of the Rashba coefficient is the same. Particularly, the maximum is located at the Al/InAs interface, precisely where the charge density is maximum [see Fig. \ref{FigB}(c,d)], what benefits the topological protection of the states populating the wire. We also note that the maximum of the SO coupling is above 10 meV$\cdot$nm for both types of crystals (even for this small $V_{\rm bg}$ value), which is large enough to create a measurable topological gap.

In our simulations we have used the parameters that correspond to (111) zinc-blende nanowires, which are the typical ones used in experiments. However, (0001) wurtzite nanowires would have provided similar features (not shown), although with smaller topological minigaps.

\section*{Simplified model}
\label{App:B}
The diagonalization of the Hamiltonian of Eq. \eqref{Eq:Hamiltonian} is computationally demanding because of the small grid spacing imposed by the large Fermi momentum of the Al layer. For this reason, we are only able to use this model to obtain the band structure and DOS for certain gate voltages, but not to compute the phase diagram of the studied devices in a broad range of parameters. To overcome this problem, we use a less demanding model that includes the induced pairing and exchange field in the wire in an heuristic way (as energy-independent but position-dependent parameters). This \textit{simplified} model only considers the wire, and thus the effective Hamiltonian can be discretized using a larger lattice spacing ($h=1$ nm in our simulations). In the following subsections we explain how both phenomena (the induced superconducting and magnetic effects) can be incorporated to reproduce qualitatively the behaviour of the full model.

\begin{figure}[h!]
\includegraphics[width=1\columnwidth]{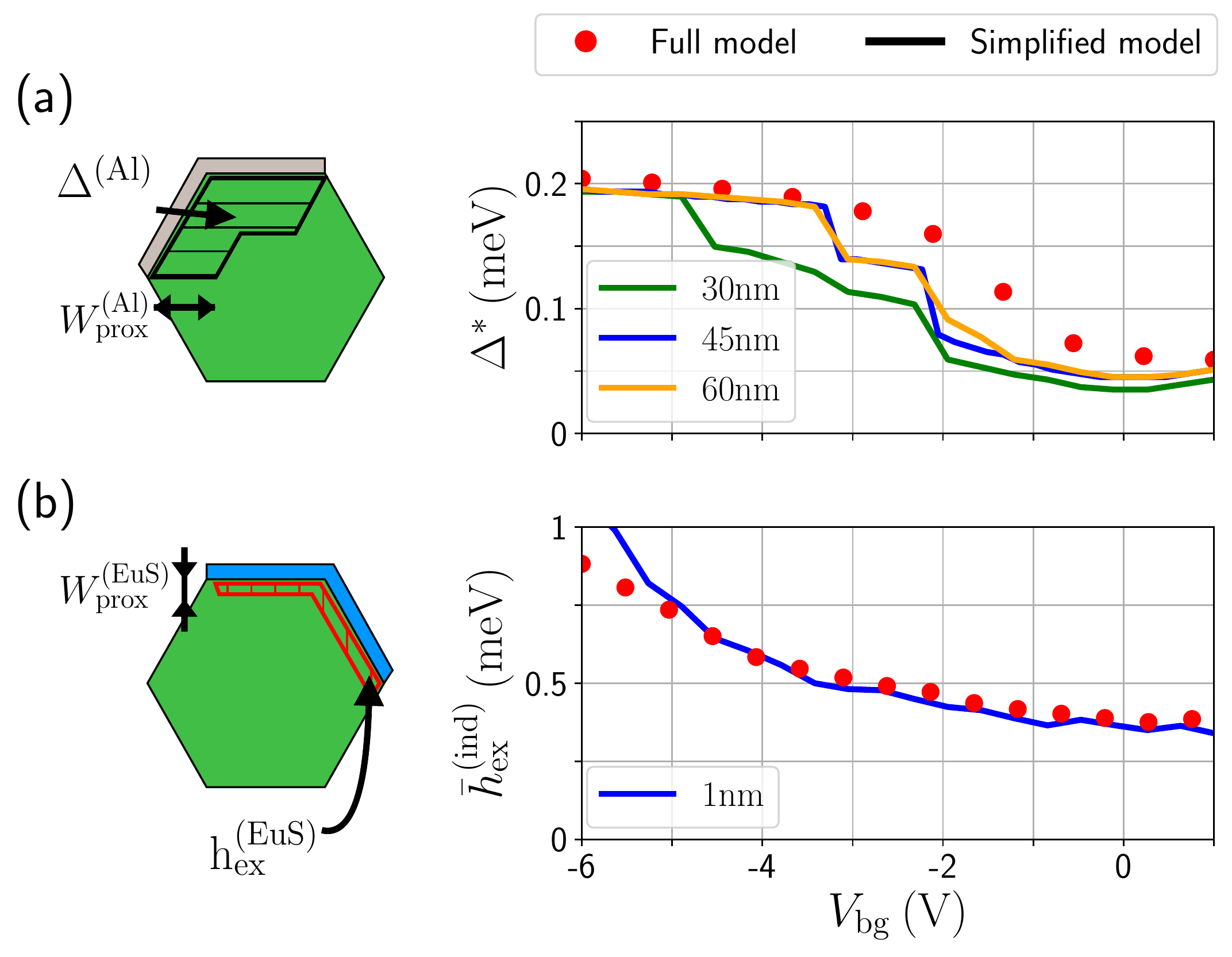}
\caption{(a) Comparison of the induced superconducting gap $\Delta^*=\mathrm{min}\left(|E(k_z)|\right)$ versus back-gate potential between the full model (dots) and the simplified model (lines). To the left, schematics of the device. Two facets of the wire are covered with an 8 nm wide Al layer. The full model includes the superconducting layer in the Hamiltonian at the same level as the nanowire. In the simplified model the Al layer is integrated out and its superconducting proximity effect on the wire Hamiltonian is included as a region of width $W_{\rm prox}^{\rm (Al)}$ with pairing amplitude $\Delta^{\rm (Al)}$ (marked as a streaked region in the schematics). To the right, $\Delta^*$ versus $V_{\rm{bg}}$; in the simplified model we take $\Delta^{\rm (Al)}=0.23$ meV and consider different $W_{\rm prox}^{\rm (Al)}$ (in colors). (b) Same as (a) but now comparing the induced exchanged field $\bar{h}_{\rm ex}^{\rm (ind)}$ versus $V_{\rm{bg}}$. The EuS shell covers two facets and has a width of 8 nm in the full model calculations. In the simplified model we take $W_{\rm prox}^{\rm (EuS)}=1$ nm with an exchange field of $h^{\rm (EuS)}_{\rm ex}=100$ meV.}
\label{FigB}
\end{figure}

\subsection*{1. Induced superconductivity}
We consider the superconducting proximity effect in the nanowire effectively by introducing a region close to the InAs/Al interface with a finite pairing amplitude $\Delta^{\rm (Al)}$ [see sketch of Fig. \ref{FigB}(a)]. The parameter $\Delta^{\rm (Al)}$ and the width of this proximitized region, $W_{\rm prox}^{\rm (Al)}$, are then chosen such that this simplified model reproduces (approximately) the same behaviour for the induced superconducting gap as the full model. 

In Fig. \ref{FigB}(a) we show the evolution of the induced gap $\Delta^*$ versus the back-gate potential. Dots correspond to calculations using the full model while solid lines correspond to the simplified one. Different colors are used for different $W_{\rm prox}^{\rm (Al)}$ widths. We find that $W_{\rm prox}^{\rm (Al)}=45$ nm is the best fit to the full model results.

\subsection*{2. Induced exchange field}
Following the same reasoning as for the induced pairing, we model the induced magnetization by introducing a region close to the InAs/EuS interface with a finite exchange field $h_{\rm ex}^{\rm (EuS)}$ [see sketch in Fig. \ref{FigB}(b)]. Once again, we choose $h_{\rm ex}^{\rm (EuS)}$ and the width of this proximitized region, $W_{\rm prox}^{\rm (EuS)}$, so as to reproduce approximately the results obtained with the full model. In Fig. \ref{FigB}(b) we show the mean induced exchange field $\bar{h}_{\rm ex}^{\rm (ind)}=\sum_j f(E^{(j)}) \left<h_{\rm ex}\sigma_x \right>_j$ versus the back-gate potential {(where $f$ is the Fermi-Dirac distribution and $j$ is summed over all populated subbands )}. Dots correspond to the full model calculations and solid lines to the simplified model using $h^{\rm (EuS)}_{\rm ex}=100$ meV and $W_{\rm prox}^{\rm (EuS)}=1$ nm. For the gate voltage range studied in this work (i.e. $V_{\rm bg}>-4$V) the simplified model provides a very good agreement with the full model.
The characteristic penetration length of electrons into the insulating EuS is given (roughly) by the magnetic length
\begin{equation}
\lambda=\left( \sqrt{\left| k_{\rm F}^2-\frac{2m_{\rm eff}E_{\rm F}^{\rm (EuS)}}{\hbar^2} \right| }\right)^{-1} \sim 1 \ \mathrm{nm},    
\end{equation}
which is the value of the proximitized width $W_{\rm prox}^{\rm (EuS)}$ that we use in our simplified model.

In addition to this \textit{direct} proximity effect from the EuS into the InAs, there is an (ungateable) induced exchange field from the EuS into the Al of the order of 0.06 meV, see full-model calculations in main text. This proximity effect is present at the interfaces where Al and EuS overlap in the overlapping device (thus, it does not appear in the non-overlapping device). This effect causes an \textit{indirect} proximity effect from the EuS into the InAs through the Al. We include this effect by adding an exchange field $h_{\rm ex}^{\rm (Al)}=0.06$ meV into the superconducting proximitized region discussed in the previous subsection. 

\begin{figure*}[h!]
\includegraphics[width=1.95\columnwidth]{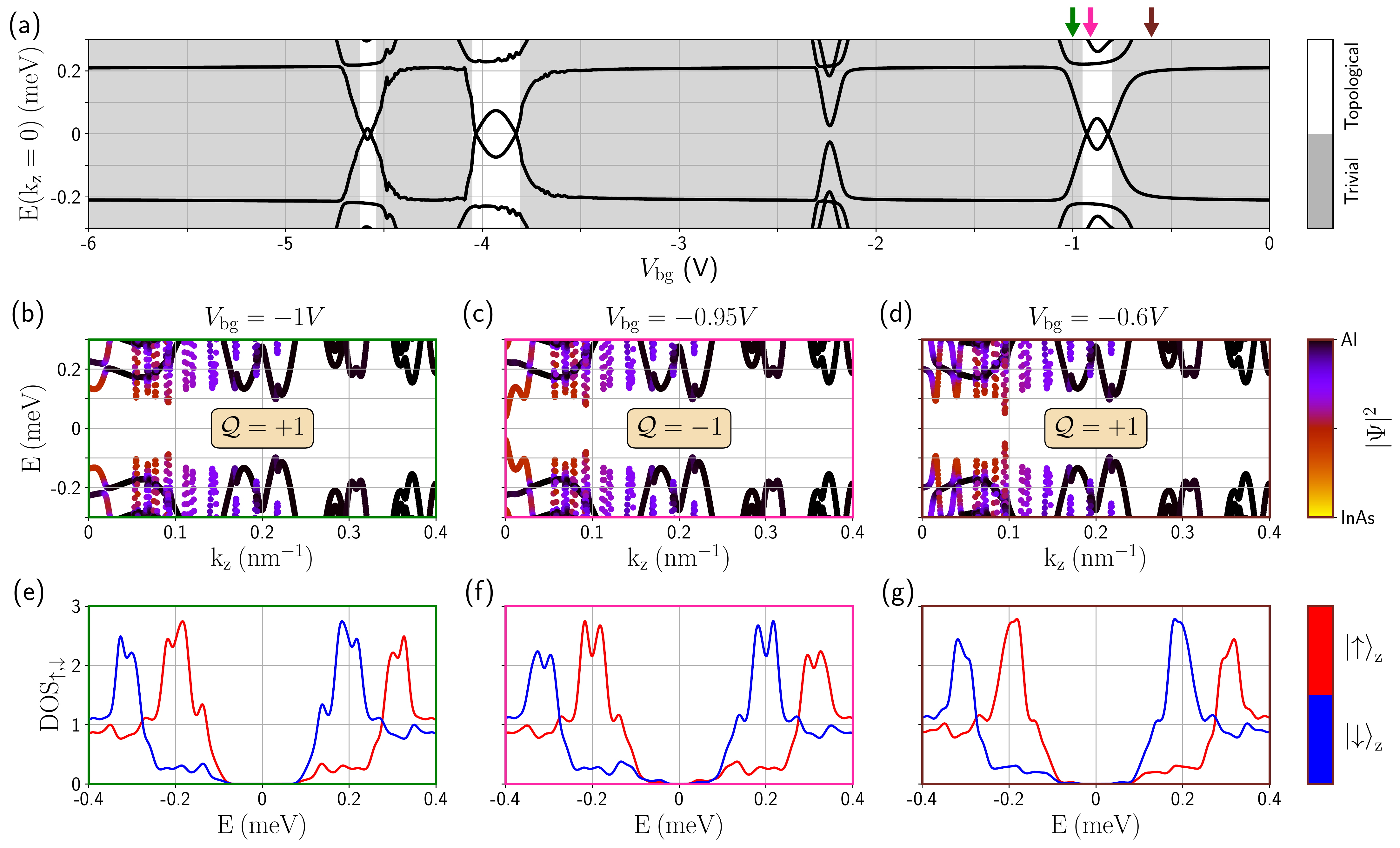}
\caption{(a) Energy spectrum at $k_z=0$ versus the back-gate voltage $V_{\rm bg}$ using the full model for the overlapping device. Grey (white) color corresponds to a trivial (non-trivial) topological phase. This can be established by computing the topological invariant $\mathcal{Q}$, as explained in the text. (b-d) Bandstructure versus $k_z$ for different gate voltages around the last topological region (see arrows): (b) before, (c) inside, and (d) after the topological region. With colored bars we show the weight of the eigenstate in the Al layer (black) and in the InAs wire (yellow). In (e-f) we show the spin-polarized DOS for the same back-gate voltages indicated with arrows in (a). We set $V_{\rm sg}^{\rm (L,R)}=0$.}
\label{FigC1}
\end{figure*}

\begin{figure*}
\includegraphics[width=1.95\columnwidth]{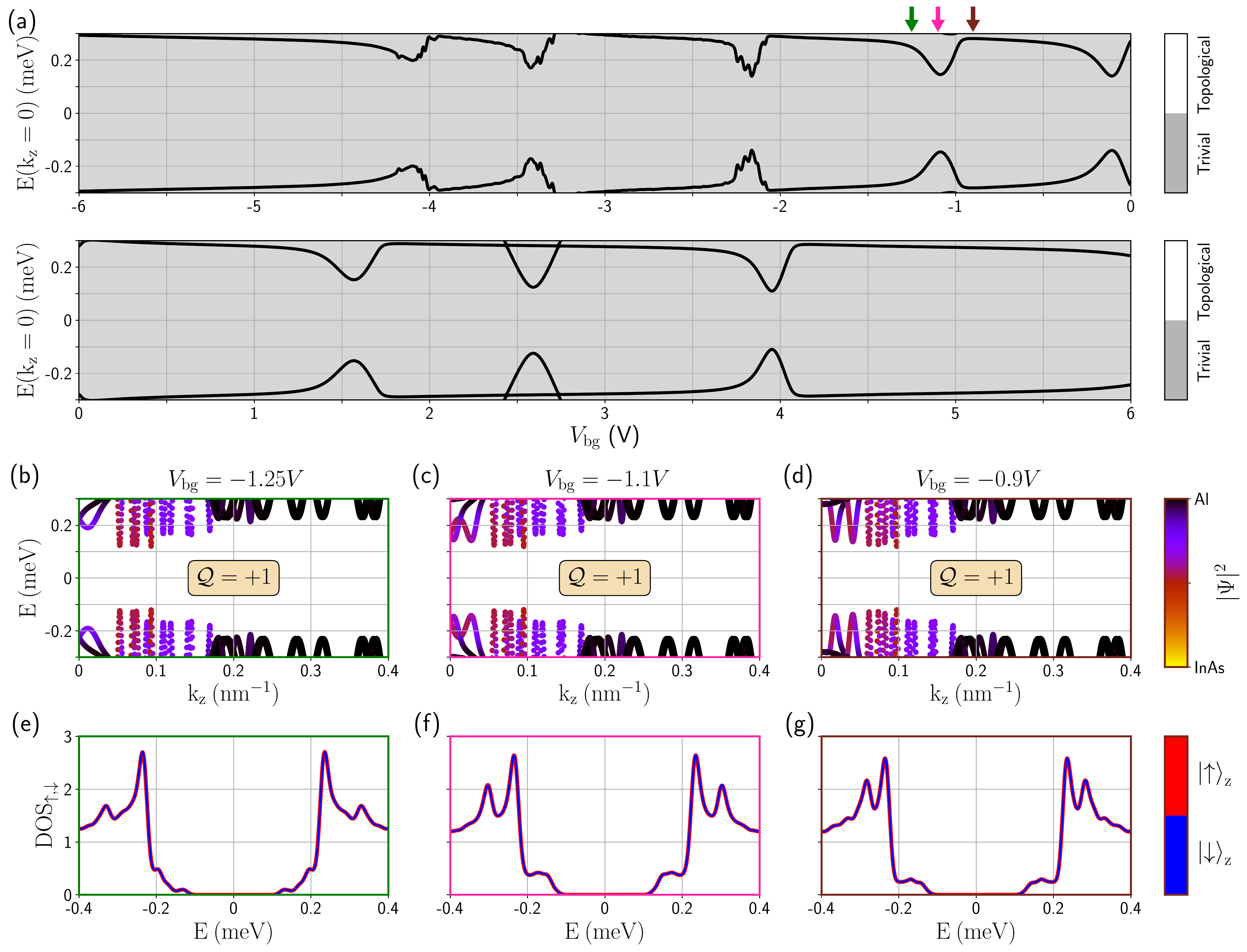}
\caption{Same as Fig. \ref{FigC1} but for the non-overlapping device.}
\label{FigC2}
\end{figure*}

\section*{Topological phase diagram for the full model}
\label{App:C}
In the main text we study the topological phase diagram (versus $V_{\rm{bg}}$ and $h_{\rm{ex}}^{(\rm{EuS})}$) using the simplified model. In this section we analyze the phase diagram versus the back gate potential using the full model, which is more computationally demanding. Note that $h_{\rm{ex}}^{(\rm{EuS})}$ is analyzed as a free parameter in the phase diagram of the simplified model, but it is a built-in quantity within the full model. This analysis complements the calculations performed in Figs. 2 and 3 of the main text.

In Fig. \ref{FigC1} we consider the case of the overlapping device. In panel (a) we show the energy spectrum at $k_z=0$ versus $V_{\rm{bg}}$. For the voltage range shown there, we observe low energy pairs of states that cross zero energy at (roughly) $V_{\rm bg}=-4.5$ V, $V_{\rm bg}=-4$ V and $V_{\rm bg}=-1$ V. These zero-energy crossings of the bulk states are related to topological phase transitions. This is demonstrated by the calculation of the topological invariant shown with white-grey colors (see the first section of this Supplemental Material for details on this calculation). In Fig. \ref{FigC1}(b-d) we show the dispersion relation versus the momentum along the wire's direction $k_z$ for three specific $V_{\rm bg}$ values: (b) at $V_{\rm bg}=-1$ V, i.e., before the last topological region; (c) at $V_{\rm bg}=-0.95$ V, inside the topological region; and (d) at $V_{\rm bg}=-0.6$ V, after this region. In the top panel of these three figures we show with colors the weight of the states in the different materials, from a state completely located in the Al layer (black) to completely located in the InAs wire (yellow). The states that cross zero energy in this topological region correspond to mixed states with half of the wavefunction located in the Al and the other half in the InAs. For completeness, we also show in Fig. \ref{FigC1}(e-f) the spin-polarized DOS for the bandstructures shown in Fig. \ref{FigC1}(b-d). 

We now perform the same analysis but for the non-overlapping device (see Fig. \ref{FigC2}). We find that there are no topological regions for any value of $V_{\rm bg}$, positive or negative, at least for the right-side gate voltage used in this simulation [see Fig. \ref{FigC2}(a)]. This is because the induced exchange field in this device is much smaller than the induced gap [see Fig. \ref{FigC2}(b-d)] and therefore no topological phase can be developed.

\begin{figure*}
\includegraphics[width=1.98\columnwidth]{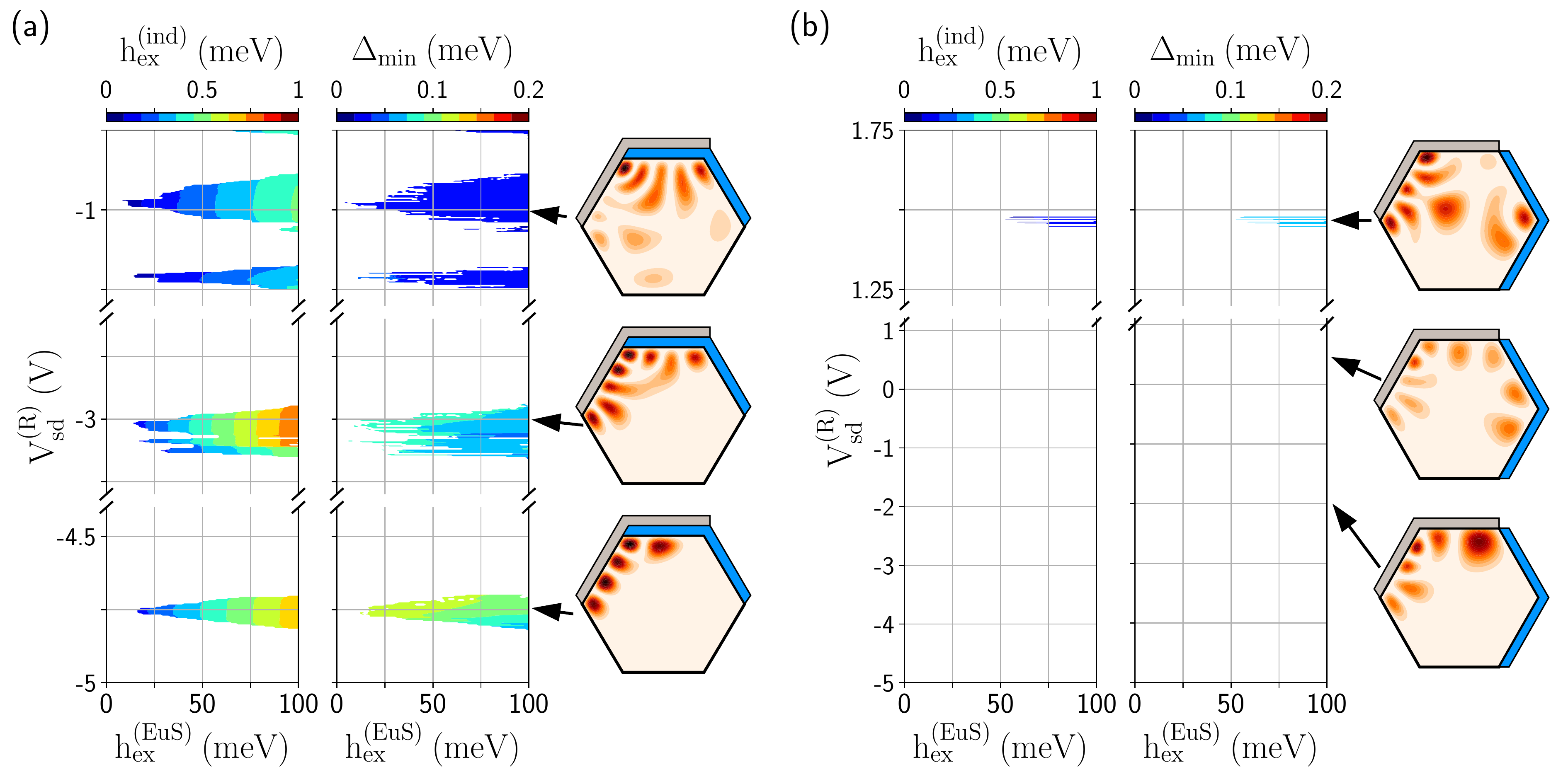}
\caption{Phase diagram of the overlapping device (a) and the non-overlapping one (b) versus the right-side gate potential $V_{\rm sg}^{\rm (R)}$ and the exchange field at the EuS-InAs interface $h_{\rm ex}^{\rm (EuS)}$, for $V_{\rm bg}=-4\mathrm{\ V}$. White means trivial (i.e., $\mathcal{Q}=+1$). With colors we show the induced exchange field in the left panel, the minigap in the middle one, and the wavefunction profile in the right one, all of them for the transverse subband closest to the Fermi energy. The $V_{\rm sg}^{\rm (R)}$ values for which the wavefunctions are plotted are pointed with arrows.}
\label{FigD}
\end{figure*}

\section*{Topological phase diagrams for the simplified model versus side-gate potential}
\label{App:D}
Using the simplified model, in the main text we study the topological phase diagram versus the back-gate potential. This gate not only tunes the chemical potential inside the wire, but it also moves upwards and downwards (along the $y$ axis) the charge probability density. This has a tremendous impact on this kind of heterostructure because, depending on the wavefunction position, the hybridization with the Al or the EuS layer dramatically differs. In the same way, the side gates play a similar role, but they change the position of the wavefunction along the perpendicular axis (the $x$ axis). Just by looking at the geometries of both devices (see Fig. 1 of the main text) one can infer that if the side gates push the wavefunction closer to the Fermi level towards the right part, the hybridization with the EuS will be increased while the hybridization with the Al will be suppressed.

To show this fact and to prove that the systems exhibit similar features in their topological phase diagrams regardless of whether the back gate or the side gates are tuned, we compute the topological phase diagram versus the right-side gate \footnote{Note that, because right and left side gates are placed parallel to each other and symmetric with respect to the wire, they provide comparable phase diagrams.}. We show the results in Fig. \ref{FigD}(a) for the overlapping device, and in Fig. \ref{FigD}(b) for the non-overlapping one. In the first two columns, we show with colors the topological regions (white means trivial, $\mathcal{Q}=1$). Particularly, colors in the first column correspond to the induced exchange field $h_{\rm ex}^{\rm(ind)}$ for the (topological) subband closest to the Fermi energy, and colors in the second column correspond to its minigap $\Delta_{\rm min}=\left|E(k=k_{\rm F})\right|$. In addition to this, we show on the right of each subfigure the probability density across the wire section of the lowest transverse subband (at $k_z=0$) at the $V_{\rm sg}^{\rm (R)}$ value pointed with the arrows.

Let us discuss first the results for the overlapping device [Fig. \ref{FigD}(a)]. This geometry exhibits several topological regions that span across potential ranges of the order of $250$ mV. Increasing the right-side gate voltage for a fixed $h_{\rm{ex}}^{(\rm{EuS})}$ (see left panel) keeps the induced exchange field almost constant, while it decreases the minigap (see right panel). This is precisely because the lowest-energy wavefunction is being pushed towards the EuS layer, moving away from the Al layer (see charge density plots on the right).

In the non-overlapping device [Fig. \ref{FigD}(b)], we observe that there is no topological phase for negative right-side gate potential values. However, there are some small topological regions for positive gate potentials. This happens because the wavefunction is located close to the EuS layer for these gate potentials (see charge density plots on the right), acquiring a non-negligible induced exchange field. But due to the large doping of the wire, these states have a large kinetic energy, what implies that they spread all across the wire. Hence, both quantities, the induced exchange field and the minigap, are small or even negligible for these states.

\section*{Two-Al-facet geometry}
\label{App:E}
In this last section we consider a different geometry than those studied in the main text. This geometry was analyzed experimentally but didn't provide zero bias peaks \cite{Vaitiekenas:NatPhys20}. Strikingly, this device is identical to the overlapping device of Fig. 1(a) of the main text, but with an additional InAs facet covered by the Al shell [see Fig. \ref{FigE}(a) for a sketch]. In Fig. \ref{FigE}(b) we show the lowest energy spectrum versus the back-gate potential at $k_z=0$. With grey/white colors we show whether the heterostructure has a trivial/non-trivial topological phase. For these gate voltages, there is only one topological region at (roughly) $-1.2$ V. However, its extension and the size of its minigap are negligibly small, in sharp contrast to the overlapping device studied before. The bandstructure around this topological region is shown in (c-e). The lowest-energy states correspond to mixed states, but mainly located in the Al. More particularly, the lowest-energy wavefunction is mainly located at the left corner of the wire (not shown), where the band-bending of the superconductor leads to a larger accumulation of charge. This large hybridization with the superconductor is what suppresses the hybridization with the EuS, leading to a small exchange field induced directly in the InAs. And this in turn is what suppresses the appearance of topological states in this device.

\begin{figure*}
\includegraphics[width=1.95\columnwidth]{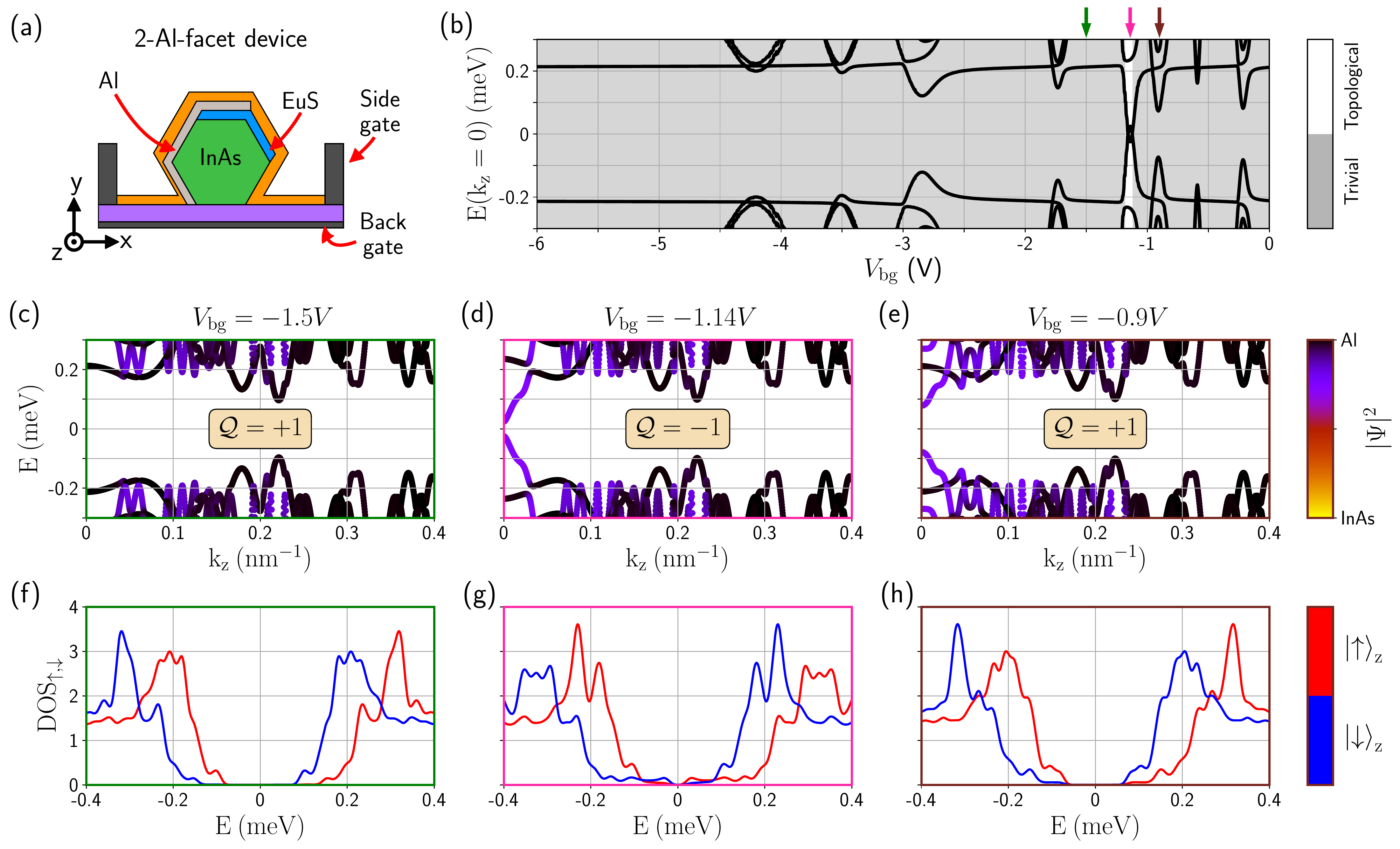}
\caption{(a) Sketch of the two-Al-facet device, which is equal to the overlapping geometry in the main text but with one more facet of the nanowire covered by the Al shell (the bottom-left one). (b) Energy spectrum at $k_z=0$ versus the back-gate voltage for this device. White/gray colors denote the non-trivial/trivial topological phases, computed with the topological invariant $\mathcal{Q}$ (see text). (c-e) Band structure versus the momentum $k_z$ for three different back-gate voltages. Their corresponding DOS are shown in (f-h). Parameters are the same as in Fig. 2 of the main text. }
\label{FigE}
\end{figure*}

\end{document}